\documentclass[fleqn,usenatbib]{mnras}
\usepackage{graphicx,pdflscape}

\def\EBV{\mbox{$E(B-V)$}}
\def\mMV{\mbox{$(m-M)_V$}}
\def\Zo{\mbox{$Z/Z_\odot$}}
\def\zs{\mbox{$Z_\odot$}}
\def\ms{\mbox{$M_\odot$}}
\def\ds{\mbox{$d_\odot$}}
\def\ls{\mbox{$L_\odot$}}
\def\mcl{\mbox{$M_{cl}$}}
\def\tA{\mbox{$t_{age}$}}
\def\dT{\mbox{$\Delta\,t$}}
\def\dZ{\mbox{$\Delta\,Z$}}
\def\CE{\mbox{$CE$}}
\def\DM{\mbox{$DM$}}
\def\mZ{\mbox{$Z$}}
\def\qA{\mbox{$k_F$}}
\def\qB{\mbox{$m_{TO}$}}
\def\x2{\mbox{$R_H$}}
\def\fH{\mbox{$[Fe/H]$}}
\def\fC{\mbox{$fitCMD$}}
\def\sR{\mbox{$S_R$}}

\title[\fC]{An efficient approach to extract parameters from star cluster CMDs: \fC}

\author[C. Bonatto]{Charles Bonatto\\
Departamento de Astronomia, Universidade Federal do Rio Grande do Sul, Av. Bento
Gon\c{c}alves 9500\\
Porto Alegre 91501-970, RS, Brazil\\ }

\begin{document}

\maketitle

\begin{abstract}

This work presents an approach (\fC) designed to obtain a comprehensive set of astrophysical 
parameters from colour-magnitude diagrams (CMDs) of star clusters. Based on initial mass function 
(IMF) properties taken from isochrones, \fC\ searches for the values of total (or cluster) stellar 
mass, age, global metallicity, foreground reddening, distance modulus, and magnitude-dependent 
photometric completeness that produce the artificial CMD that best reproduces the observed one; 
photometric scatter is also taken into account in the artificial CMDs. Inclusion of photometric 
completeness proves to be an important feature of \fC, something that becomes apparent especially 
when luminosity functions are considered. These parameters are used to build a synthetic CMD that 
also includes photometric scatter. Residual minimization between the observed and synthetic CMDs 
leads to the best-fit parameters. When tested against artificial star clusters, \fC\ shows to be 
efficient both in terms of computational time and ability to recover the input values.

\end{abstract}


\section{Introduction}
\label{intro}

A colour-magnitude diagram (CMD) can be considered as the observational counterpart - on a 
purely photometric parameter space - of a stellar population (of given age and metallicity) 
whose individual masses are distributed according to an initial mass function (IMF). Over 
time, the mass-dependent stellar evolution changes the colour and magnitude of each star, 
thus leading to variations in the evolutionary sequences morphology. This reasoning implies 
that the CMD morphology encapsulates fundamental properties related to the star cluster 
itself, such as the stellar mass, age, metallicity, distance and reddening, among others. 

The derivation of fundamental parameters of star clusters - in different dynamical states 
and formed in different environments - is important for a wide variety of studies,
ranging from determination of the star-formation rate in the Galaxy (e.g. \citealt{LG06}; 
\citealt{SFR}) to investigations on the {\em infant mortality} (e.g. \citealt{LL2003}; 
\citealt{GoBa06}) and the dynamical state and cluster dissolution time scales (e.g. 
\citealt{GoodW09}; \citealt{Lamers10}), among others. Broadly speaking, photometry is the 
easiest - and cheapest - way to collect data on substantial fractions of a star 
cluster's population. Transported to CMDs, the collected data then can be used to try and
characterize the stellar population by extracting (some of) its fundamental parameters. 

However, the combination of limited photometric depth with distances in excess of a few Kpc 
usually prevents the fainter stars from being detected by most of the available large-scale 
photometric surveys, thus producing partially-sampled CMDs that, in turn, may lead to a 
deficient derivation of cluster parameters. Besides, the presence of unknown fractions of 
binaries, differential reddening (especially for the very young clusters), and photometric 
scatter and completeness, also add complications to the task of using CMDs to derive 
cluster parameters. In particular, while age, distance and reddening can be relatively easy 
to estimate even in CMDs that are somewhat noisy and do not contain the fainter stars, metallicity, 
and especially mass, remain way more elusive to measure.  

Considering the potentially large scope that well-determined cluster parameters may have
in the current Galactic astrophysics, several approaches have already been developed in order to 
circumvent the difficulties associated to the process of extracting fundamental parameters from 
CMDs, e.g. \citet{NJ06}, \citet{daRio2010}, \citet{StHo2011}, and \citet{BoLiBi2012}. A summary  
of these approaches can be found in \citet{BoLiBi2012}. These approaches have been designed to 
deal with particular types of clusters and/or parameters, and some are not practical in terms of 
the required computational time. 

The approach presented in \citet{BoLiBi2012} included most of the relevant parameters (mass, age, 
star-formation spread, distance modulus, foreground and differential reddening, and binaries) that 
are expected to make up the CMDs of young clusters. In short, the idea was to find the parameters 
that produce a CMD that best reproduces the observed one. This was done by minimizing the residuals 
between the observed and simulated CMDs by means of a global optimization method that varied the 
parameters along the direction of the minimization. Although the approach was able to recover the 
input parameters for simulated CMDs, it required extremely long computational times when applied to 
actual star clusters. To minimise stochasticity, the approach worked with the average CMD of several 
mock clusters, each sharing the same set of parameters. Thus, the number of stars turned out to be 
very large, which in turn affected critically the computational time.

Based on lessons learned especially in \citet{BoLiBi2012}, a more efficient and comprehensive
approach - \fC\ - 
to extract parameters from CMDs is here presented. Instead of working with mock star clusters, it
is based on a mass function (of a given age, metallicity, and mass) displaced some distance from
the Sun, and affected by reddening and photometric scatter. An additional feature is that now 
photometric completeness is also taken into account. It is shown to be efficient both in 
terms of computational time and recovery of input values from simulated CMDs. Similarly to
\citet{BoLiBi2012}, the rationale involves the minimization of residuals between the simulated and
observed Hess diagrams (\citealt{Hess24}). 

This paper is organised as follows: Sect.~\ref{TOM} describes the approach; Sect.~\ref{TestCMD} 
discusses the efficiency of \fC\ in retrieving parameters from simulated CMDs; Sect.~\ref{APP} 
discusses the application of \fC\ to actual CMDs of open clusters, a dwarf galaxy, and a 
globular cluster. Concluding remarks are given in Sect.~\ref{CONC}.

\section{The approach: \fC}
\label{TOM}

The basic goal of \fC\ is to extract fundamental parameters of a star cluster by means of 
the photometric information contained in its observed CMD. The underlying principle is to find an 
artificial CMD that best reproduces the observed one; the set of input parameters is then assumed 
to be representative of the star cluster itself. The parameters included here are the total mass 
stored in stars (or cluster mass, \mcl), age (\tA), global metallicity (\mZ), foreground reddening 
(or colour excess, \CE) and the apparent distance modulus (\DM). Photometric completeness is also 
taken into account by means of a Fermi function. Completeness as a function of the magnitude $m$ 
is described as $f_C(m) = 1/(1+\exp{[\qA(m-\qB)]})$, where \qB\ is the turnover magnitude 
$(f_C(\qB) = 0.5)$ and \qA\ controls the steepness of the descent. Strictly speaking, this formulation
refers to the completeness (at magnitude $m$) relative to that of the brightest stars. The
analytical shape of the Fermi function appears to be convenient to describe the photometric completeness
because it changes rather slowly at the bright end, and falls off exponentially after some fainter 
magnitude. When completeness effects are irrelevant, the stellar density over the evolutionary sequences
of a stellar population on a CMD should follow approximately that of a mass function. However, completeness 
artificially decreases the stellar density towards fainter magnitudes, and the comparison with the 
intrinsic density allows fitCMD to derive the completeness function parameters \qA\ and \qB.

If binaries and differential reddening are ignored, an artificial star cluster CMD can be built by
defining the following minimum set of parameters: \mcl, \tA, \mZ, \CE\ and \DM. Individual stellar 
masses are attributed according to an initial mass function (IMF), which can be chosen either as 
that of \citet{Salpeter55} or the segmented distribution of \citet{Kroupa01}. Photometric scatter 
must also be added for more realism. In addition, a mass to light relation (MLR) - usually taken from
isochrones - is required to produce the CMD representation of a mock star cluster. Examples of such
experiments are in, e.g. \citet{BoLiBi2012}.

Regarding isochrones, this work employs the latest PARSEC\footnote{Downloadable from 
$http://stev.oapd.inaf.it/cgi-bin/cmd$} v1.2S$+$COLIBRI PR16 (\citealt{Bressan2012};
\citealt{Marigo2017}) models, although any other system having an adequate coverage in age, 
metallicity and stellar mass can be used. The present work employs filters that are part of the 
following photometric systems {\em (i)} \citet{Bessell90} \& \citet{BesBrett88} (filters B, V, J, 
and K), {\em (ii)} DECAM (filters $g$ and $r$), and  {\em (iii)} WFC/ACS (filters F606W and F814W). 
These isochrones contain stars with masses larger than $0.1\ms$ and are computed for a scaled-solar 
composition, and follow the relation $Y=0.2485+1.78Z$, where Y is the He content and Z is the 
total metallicity; the solar metal content is $\zs=0.0152$. They also include the pre-main sequence 
phase. For an adequate coverage in age, the following values are considered: $1-10$\,Myr (in steps 
of $\dT=1$\,Myr), $10-20$\,Myr ($\dT=2$\,Myr), $20-50$\,Myr ($\dT=5$\,Myr), $50-100$\,Myr ($\dT=10$\,Myr), 
$100-500$\,Myr ($\dT=25$\,Myr), $500-1000$\,Myr ($\dT=50$\,Myr), $1000-13500$\,Myr ($\dT=250$\,Myr). 
In terms of the metallicity \mZ, the sampling is: from $10^{-4}-10^{-3}$ (steps of $\dZ=10^{-4}$), 
and $10^{-3}-3\times10^{-2}$ ($\dZ=10^{-3}$). As a consequence of such a relatively high resolution 
in age and metallicity, the total number of isochrones is 3978, but smaller ranges can be considered 
as well (see below).

Consider an artificial star cluster ($SC_O$) characterized by the set of intrinsic parameters 
$(\mcl,\tA,Z,IMF)$. Obviously, any other model star cluster built with the same parameters will 
be a twin representation of $SC_O$. But, given the statistical nature of the individual stellar 
masses distribution - and, to some extent, their colours and magnitudes, the corresponding CMD may 
end up presenting significant differences with respect to that of $SC_O$, especially for young and 
low-mass cases (e.g. \citealt{BoLiBi2012}). However, the fundamental point here is that the IMF is 
the same in all cases. So, instead of the discrete - and stochastic - CMDs, the present approach is 
based on properties of the IMF itself; a brief description is sketched below.

\begin{figure}
\begin{minipage}[b]{0.5\linewidth}
\includegraphics[width=\textwidth]{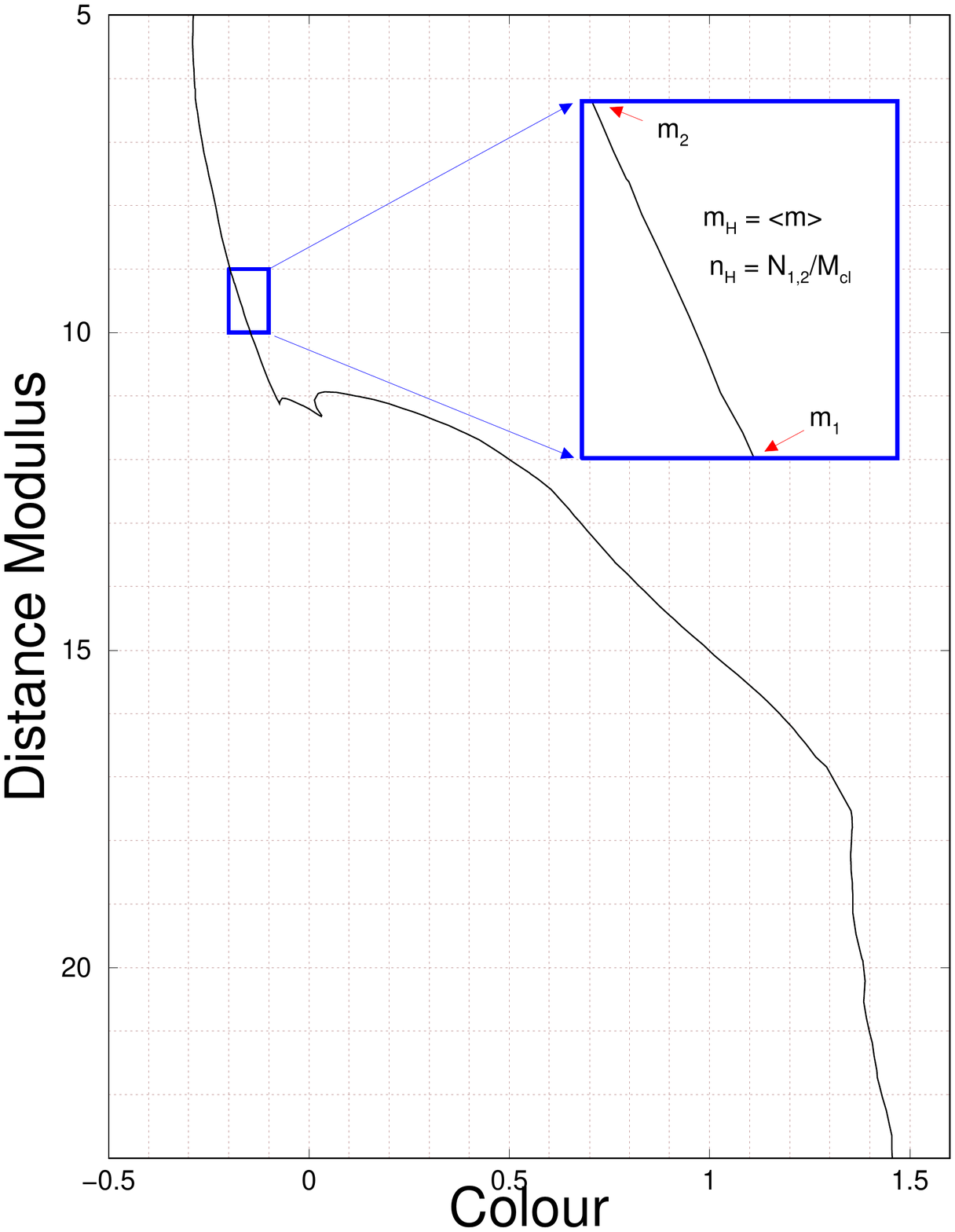}
\end{minipage}\hfill
\begin{minipage}[b]{0.5\linewidth}
\includegraphics[width=\textwidth]{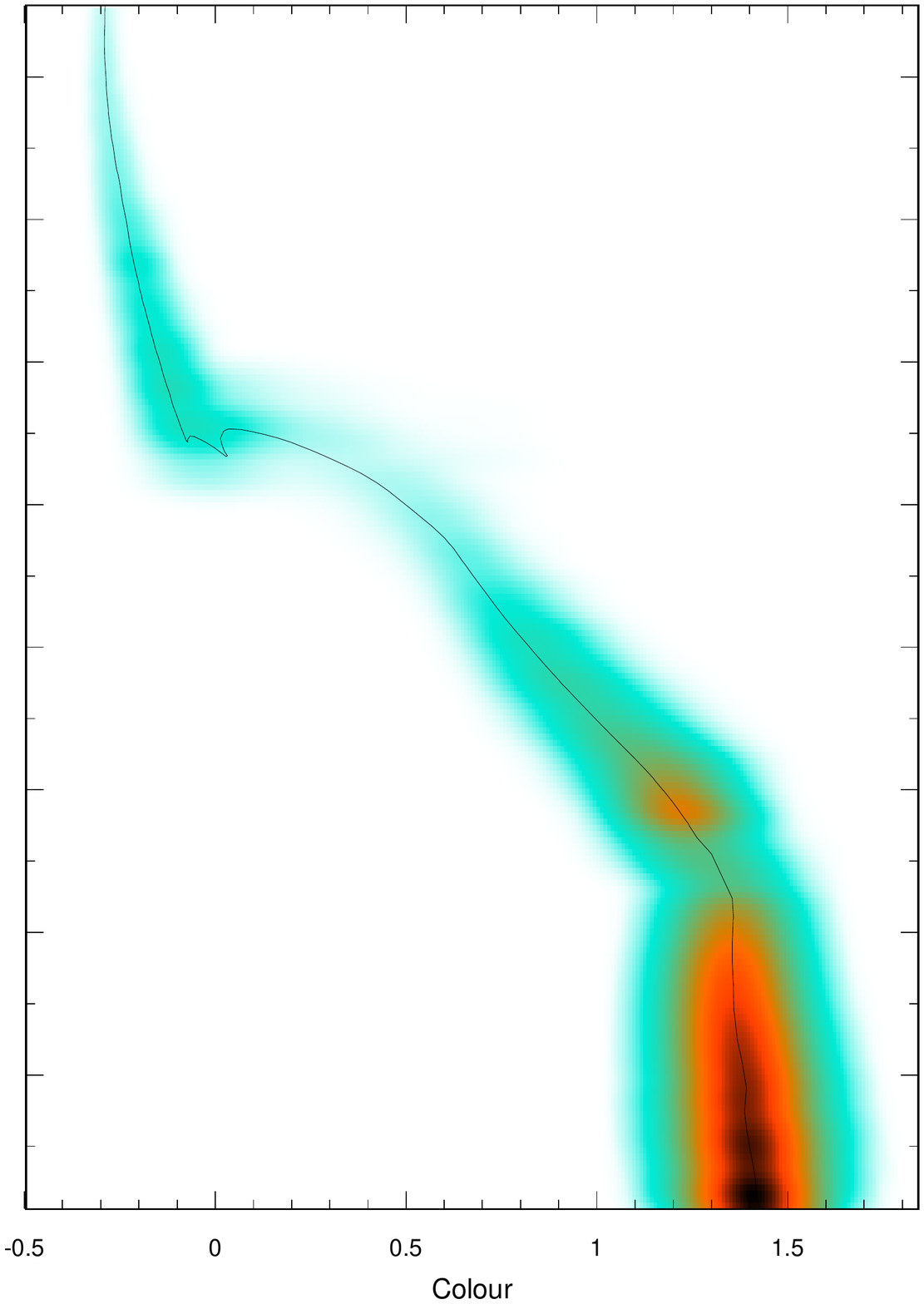}
\end{minipage}\hfill
\caption[]{Left panel: given an IMF (represented by an isochrone), a CMD cell will contain 
$N_{1,2}$ stars with mass in the range $(m1,m2)$, with an average stellar mass $<m>$; the 
stellar density is defined as $n_H=N_{1,2}/\mcl$. Right: Hess diagram - with photometric
scatter added - of the CMD at left.}
\label{fig1}
\end{figure}

The first step consists in defining the ranges in which the parameters (\mcl, \tA, \mZ, \CE, \DM) 
will be searched. This, in turn, also sets the actual number of isochrones to be dealt with. With 
the IMF, a CMD is built for each isochrone (at $\DM=0$ and no reddening), with a discretization 
in the colour/magnitude plane also defined at the beginning. So, a given CMD cell - containing stars 
with mass in the range $(m1,m2)$) - will have the number of stars per cluster mass $n_H=N_{1,2}/\mcl$,
as well as the average mass ($<m>$), which basically involves integrating the mass function 
($\phi(m) = dN/dm$) between $(m1,m2)$: $N_{1,2} = \int_{m1}^{m2}\phi(m)~dm$. This operation is done
a single time and before the parameter search, which drastically reduces the computational time.
After these procedures, each CMD cell will contain the respective relative density (number per 
cluster mass) of occurrence of stars, which is equivalent to the classical Hess diagram; hereafter, 
these diagrams are denoted by $H_M = H_M(\mcl,\tA,\mZ)$. This process is illustrated in Fig.~\ref{fig1}. 
The same discretization is then applied to the observed star cluster CMD when building the respective 
Hess diagram. Photometric uncertainties are explicitly taken into account in the observed Hess 
diagram (e.g. \citealt{BoLiBi2012}).

\begin{figure}
\begin{minipage}[b]{0.5\linewidth}
\includegraphics[width=\textwidth]{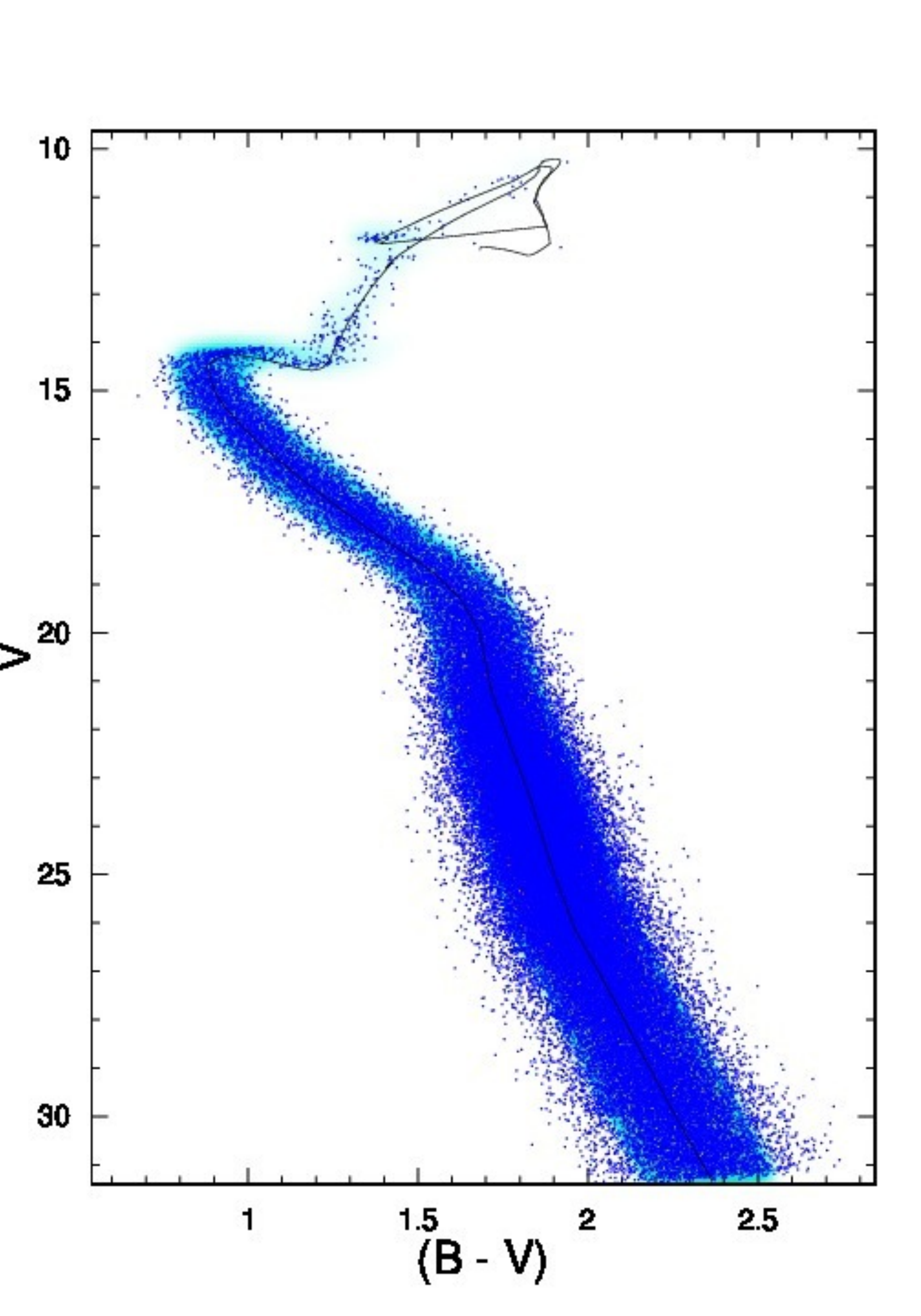}
\end{minipage}\hfill
\begin{minipage}[b]{0.5\linewidth}
\includegraphics[width=\textwidth]{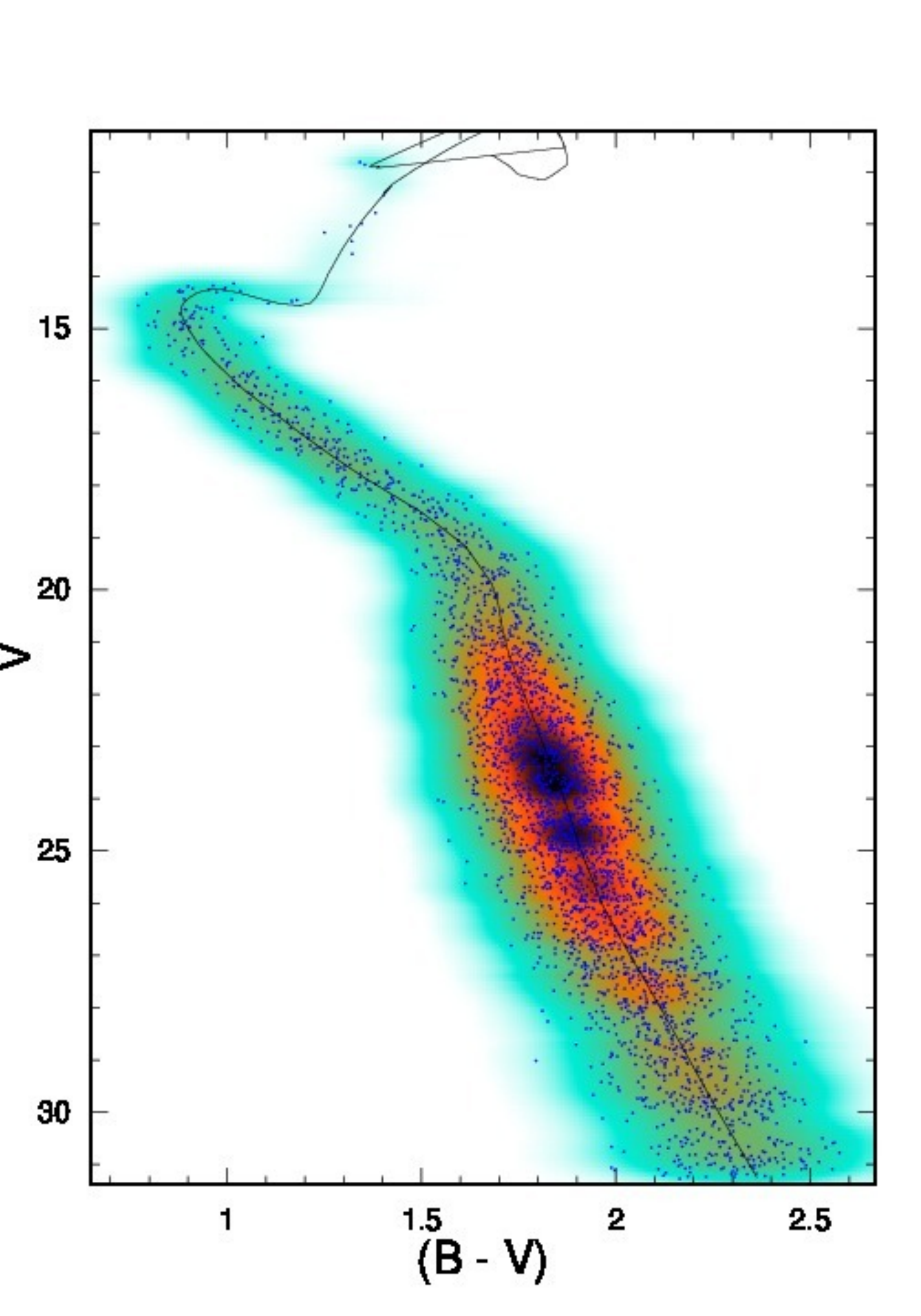}
\end{minipage}\hfill
\begin{minipage}[b]{0.5\linewidth}
\includegraphics[width=\textwidth]{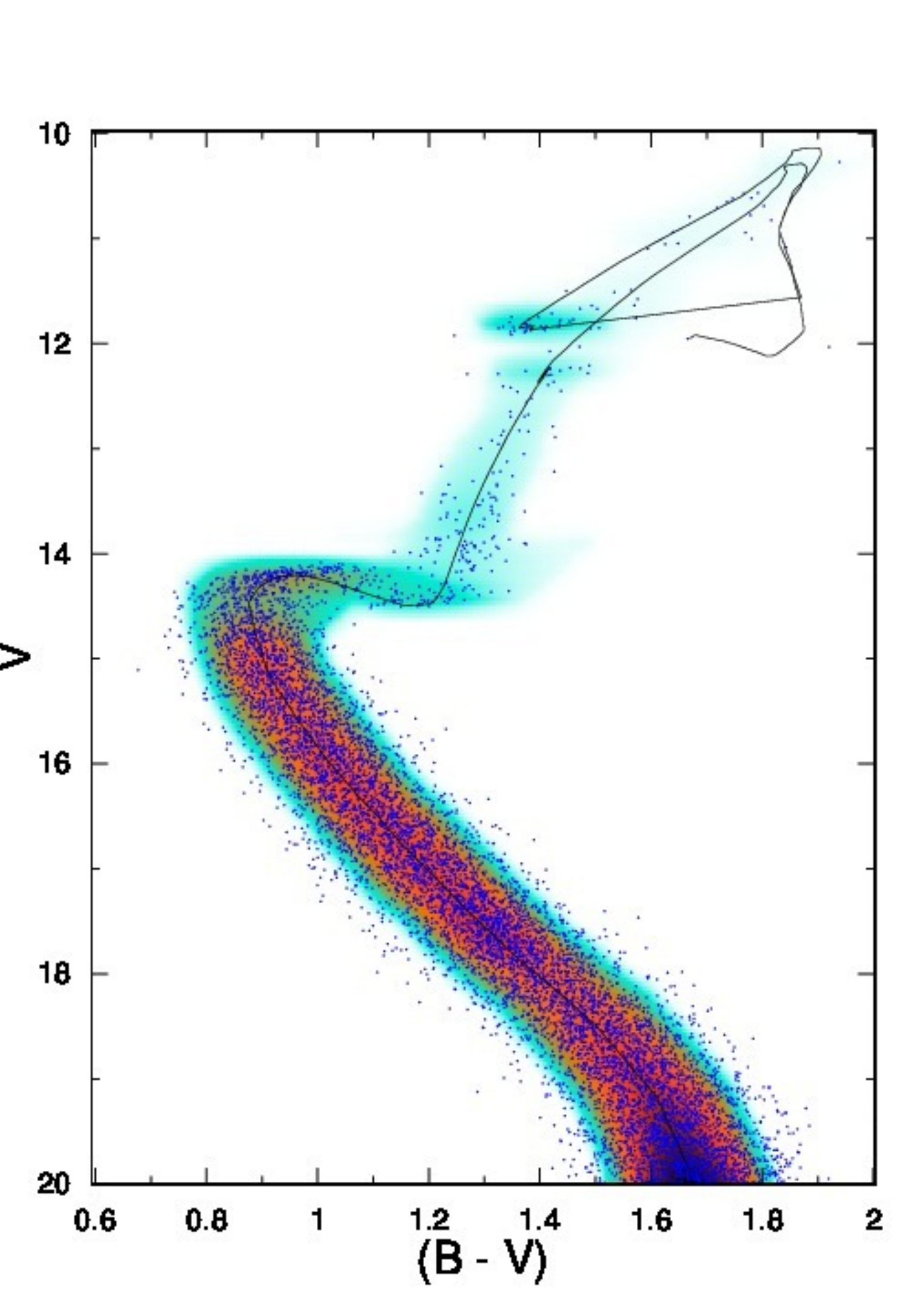}
\end{minipage}\hfill
\begin{minipage}[b]{0.5\linewidth}
\includegraphics[width=\textwidth]{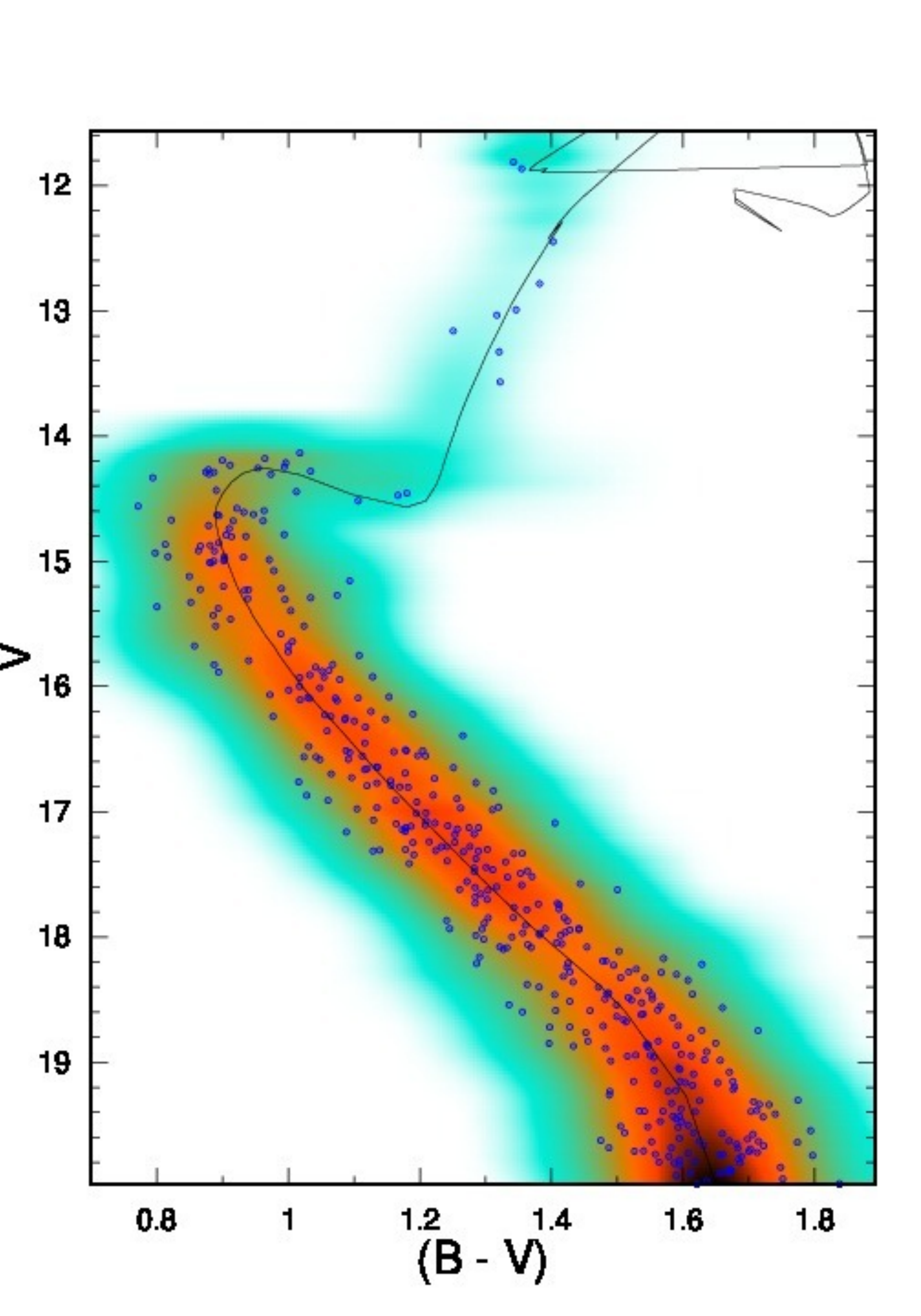}
\end{minipage}\hfill
\begin{minipage}[b]{0.5\linewidth}
\includegraphics[width=\textwidth]{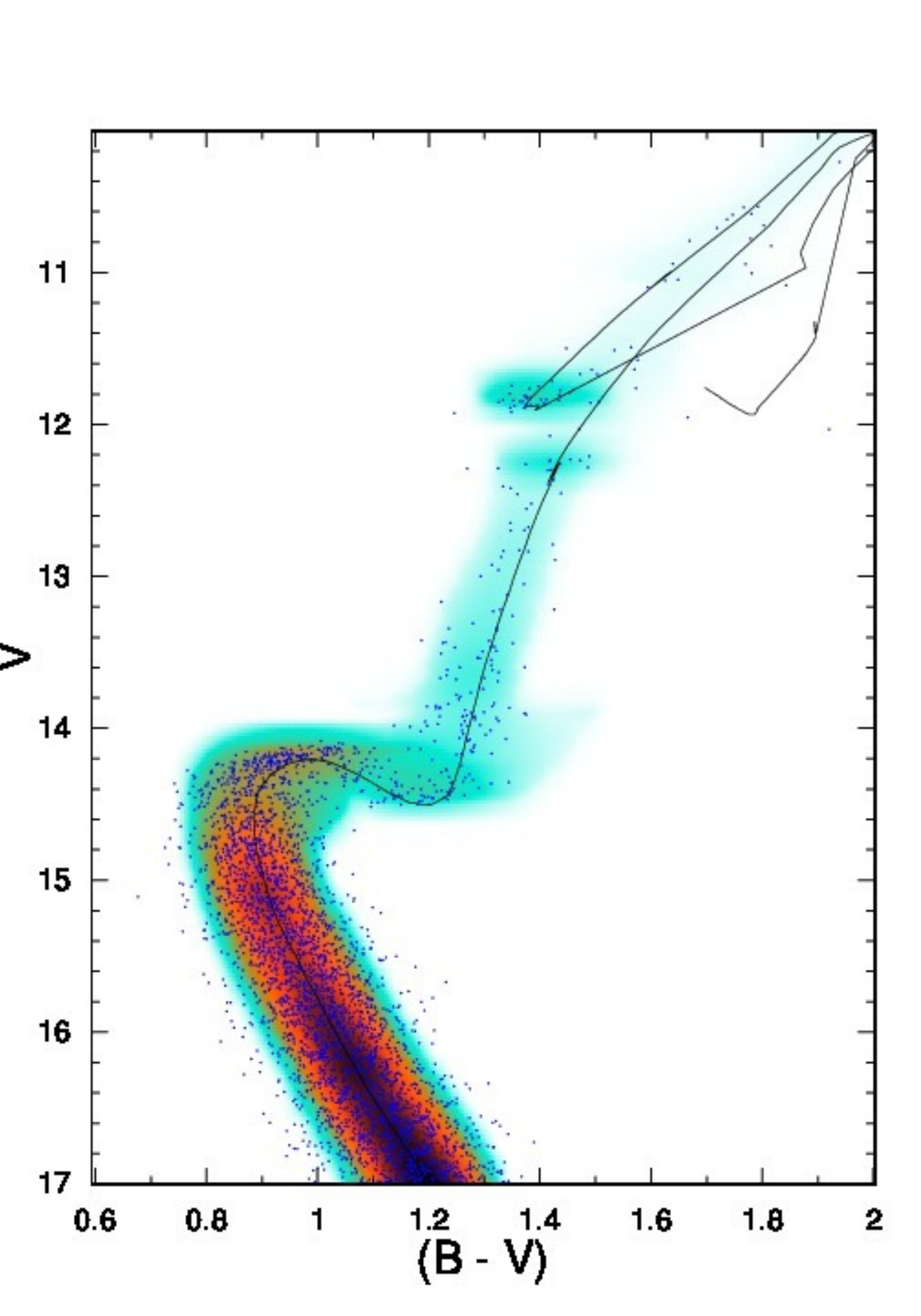}
\end{minipage}\hfill
\begin{minipage}[b]{0.5\linewidth}
\includegraphics[width=\textwidth]{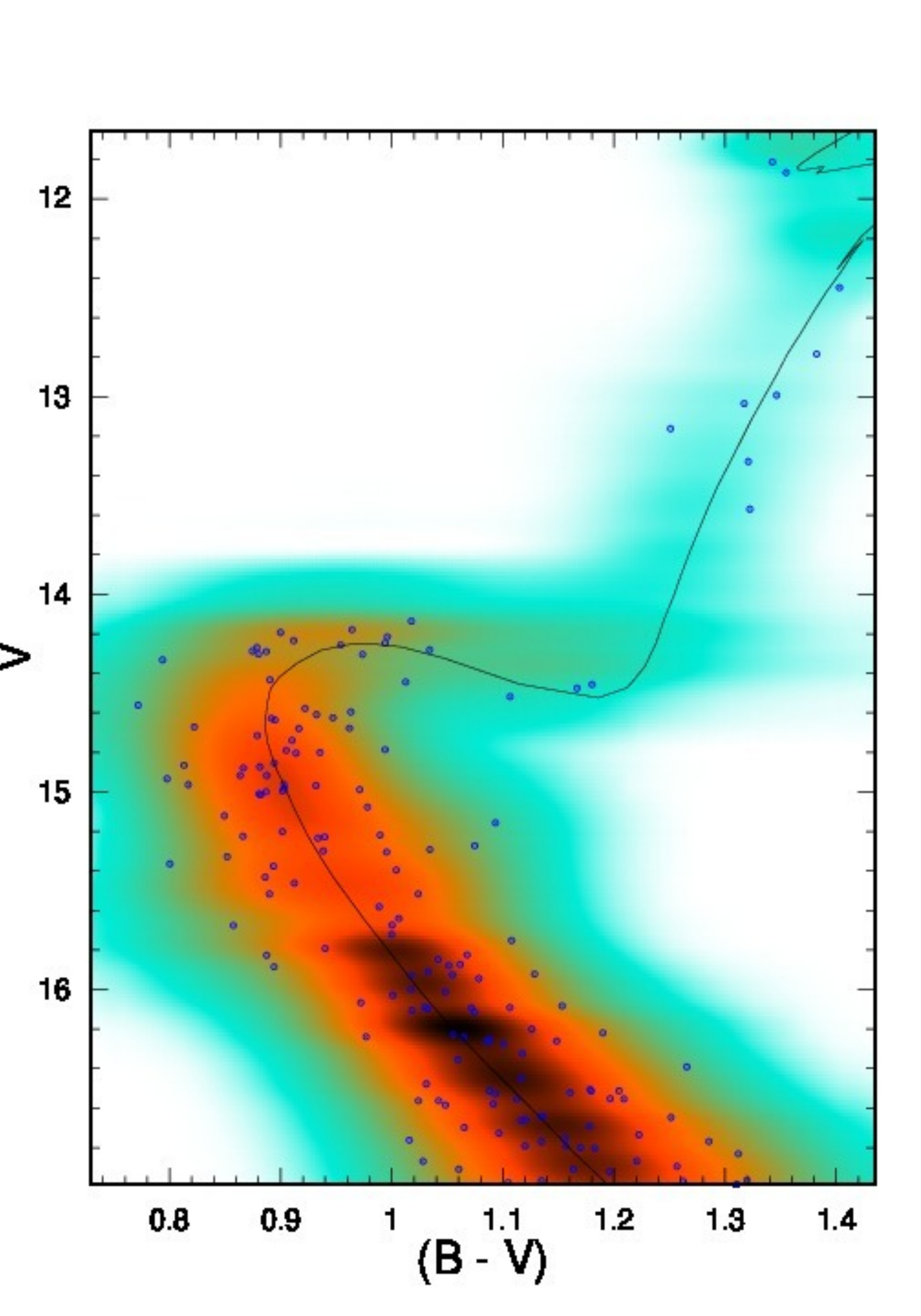}
\end{minipage}\hfill
\caption[]{CMDs (blue dots) of simulated star clusters with $\mcl=2\times10^4$\,\ms\ (left panels) 
and $\mcl=1\times10^3$\,\ms\ (right). They are analysed by \fC\ over the full magnitude range (top
panels), and partial ranges (middle and bottom). The best solutions are shown as Hess diagrams, in
which the corresponding isochrone is set.}
\label{fig2}
\end{figure}

Photometric scatter is another key component of any artificial CMD that is expected to reproduce 
that of a star cluster. Consider the rows corresponding to the magnitude in the observed Hess diagram. 
For each row we compute the total stellar density and build the corresponding distribution function 
as a function of colour (columns). To minimize local fluctuations, the stellar density function for 
row $N$ is actually the average of those in rows $N-1$, $N$ and $N+1$. After identifying the colour 
where the maximum occurs, we compute the fraction of the total stellar density that each colour cell 
contains, both for bluer and redder colours with respect to the maximum. This broadening function is 
then applied to the respective row (magnitude value) in the simulated Hess diagram (see below). Thus, 
\fC\ allows both for colour-asymmetric and magnitude-dependent scatter. Comparisons of the 
actual photometric scatter with the simulated ones can be seen in Figs.~\ref{fig2}-\ref{fig4}.

\begin{table*}
\caption[]{Recovery of model parameters}
\label{tab1}
\renewcommand{\tabcolsep}{5.2mm}
\begin{tabular}{cccccccc}
\hline\hline
Range&N&\mcl&\tA&\mMV&\EBV&\Zo&$M_{CMD}$\\
(mag)& (stars) &  (\ms)  &(Gyr)&(mag)& (mag)& &(\ms)  \\
(1)&(2)&(3)&(4)&(5)&(6)&(7)&(8)\\
\hline
Model&61034&20000&5&10.94&0.30&1.18&20000\\
\hline
Full &61034&$20000^{+642}_{-598}$&$5.0^{+0.3}_{-0.4}$&$10.94^{+0.03}_{-0.03}$&$0.30^{+0.01}_{-0.01}$&$1.18^{+0.02}_{-0.04}$&20000\\
$V<20$ &8643&$20000^{+1870}_{-1590}$&$5.0^{+0.24}_{-0.17}$&$10.94^{+0.03}_{-0.02}$&$0.30^{+0.01}_{-0.0}$&$1.18^{+0.04}_{-0.03}$&7400\\
$V<17$ &3630&$20500^{+2640}_{-2300}$&$5.2^{+0.2}_{-0.1}$&$10.84^{+0.03}_{-0.01}$&$0.27^{+0.01}_{-0.01}$&$1.32^{+0.05}_{-0.02}$&3900\\
\hline
Model&3103&1000&5&10.94&0.30&1.18&1000\\
\hline
Full &3103&$993^{+98}_{-84}$&$4.8^{+1.0}_{-1.1}$&$10.89^{+0.09}_{-0.07}$&$0.30^{+0.01}_{-0.01}$&$1.25^{+0.07}_{-0.07}$&990\\
$V<20$ &416&$991^{+184}_{-145}$&$4.8^{+0.4}_{-0.8}$&$10.93^{+0.05}_{-0.06}$&$0.30^{+0.01}_{-0.01}$&$1.18^{+0.12}_{-0.06}$&360\\
$V<17$ &184&$1010^{+244}_{-132}$&$5.0^{+0.2}_{-0.2}$&$10.94^{+0.06}_{-0.06}$&$0.31^{+0.02}_{-0.01}$&$1.05^{+0.04}_{-0.02}$&190\\
\hline
\end{tabular}
\begin{list}{Table Notes.}
\item Col.~(1) - magnitude threshold considered; (2) - number of stars present in the CMD; 
(3) - derived cluster mass; (4) - age; (5) - distance modulus; (6) - colour excess; (7) - 
metallicity; (8) - stellar mass actually present in CMD.
\end{list}
\end{table*}

The parameter search consists on finding the absolute minimum of the residual hyper-surface (\x2) 
defined as $\x2=\left(H_{obs}-H_{sim}\right)^2,$ where $H_{obs}$ is the observed Hess diagram and 
$H_{sim}=H_{sim}(\mcl,\tA,\mZ,\CE,\DM,\qA,\qB)$ is the simulated one. The values of \mcl, \tA, \mZ, 
\CE, \DM, \qA, and \qB\ at the absolute minimum are assumed to represent those of the star cluster. In 
practical terms, locating the minima of \x2\ is equivalent to finding the parameters that minimize the 
quantity $$\sR = \sum_{mag,col}W(mag)\times\x2(mag,col),$$ on the colour/magnitude plane. The sum runs 
over all Hess cells and $W(mag)$ is the statistical weight of each cell. $W(mag)$ corresponds to the 
inverse of the observed Hess density computed at the respective magnitude of each cell, i.e. 
$W(mag)=1/\sum_{col} H_{obs}(mag,col)$.

We employ the global optimisation method known as Simulated Annealing (SA, \citealt{SA94}). Simulated 
annealing originates from the metallurgical process by which the controlled heating and cooling of a 
material is used to increase the size of its crystals and reduce their defects. If an atom is stuck 
to a local minimum of the internal energy, heating forces it to randomly wander through higher energy 
states. In the present context, a state is the surface $\x2=\x2(\mcl,\tA,\mZ,\CE,\DM,\qA,\qB)$, 
corresponding to a specific set of values of the parameters being optimised. The slow cooling 
increases the probability of finding states of lower energy than the initial one. SA is a global 
optimisation technique that can escape from local minima (\citealt{SA94}; \citealt{BoLiBi2012}).

SA is an iterative and statistical technique that, at any given step $k$, randomly selects a new set 
of parameters ($M^k_{cl}$, $t_{age}^k$, $\mZ^k$, $\CE^k$, $\DM^k$, $k^k_F$, $m^k_{TO}$) from the respective 
ranges. This also means that the corresponding diagram $H_M^k = H_M(M^k_{cl},t_{age}^k,\mZ^k)$ will be 
the only one to be used in this particular iteration. The cells of $H_M^k$ are then multiplied by $M^k_{cl}$ 
(to end up containing the number density of stars - same as $H_{obs}$), and shifted by the values of 
$\DM^k$\ and $\CE^k$. Finally, the artificial stellar density of each cell is decreased according to the 
photometric completeness of the corresponding magnitude, and photometric scatter is added (see above). 
What results is the diagram $H_{sim}$, from which \x2\ is built. As $\sR$ decreases - i.e., a new minimum 
is found, SA concentrates on smaller parameter ranges, centred around the most promising values, and a new 
step ($k+1$) is taken. Iterations stop when SA meets the convergence criterion: 5 consecutive repetitions 
of the same value of \sR. Typical runtimes are around 1\,min on an {\em Intel Core i7 920@2.67 GHz} 
processor. However, given the statistical nature of SA, \fC\ should be repeated a few times to 
minimize the probability of getting stuck into a deep, but secondary minimum.

An interesting aspect of \fC\ is that, if the mass distribution of a cluster follows relatively 
closely a MF, the cluster mass should be retrievable quite well even when working with CMDs not 
severely truncated by limited photometric depth. In other words, if the age, metallicity, distance 
modulus and colour excess are reasonably well determined, the ratio observed/artificial Hess cells 
content - in any magnitude range - corresponds to \mcl.

\section{Testing \fC\ on artificial CMDs}
\label{TestCMD}

The efficiency of \fC\ in recovering input parameters is tested with simulated CMDs, built with 
pre-defined values of \mcl, \tA, \mZ, \CE, and \DM. Individual stellar masses are attributed 
according to Kroupa's IMF (\citealt{Kroupa01}) - for masses larger than $0.1\,\ms$ - until the sum 
matches \mcl; the respective magnitudes are taken from the PARSEC isochrone corresponding to \tA\ 
and \mZ. Typical photometric uncertainties (for the simulated distance from the Sun) are added for 
each band $k$ ($\sigma_k$). For more realism, the final photometric values for each star in band $k$ 
correspond to those in the isochrone ($m_k$) plus a displacement that is taken from a Gaussian 
distribution centred on $m_k$ and having the standard deviation $\sigma_k$, thus emulating 
observational photometric scatter. 

Colour-magnitude diagrams dealt with here correspond to the classical $V\times(B-V)$. For simplicity, 
we consider 2 models with identical age ($\tA=5$\,Gyr), apparent distance modulus ($\mMV=10.94$), colour 
excess ($\EBV=0.3$) - thus corresponding to $\ds=1$\,kpc, and global metallicity ($\Zo=1.18$), but with 
very different masses ($\mcl=2\times10^4$\ms\ and $\mcl=1\times10^3$\ms) and, consequently, number of
stars present in the CMD. Effects of photometric depth on the recovered parameters are also considered 
(Table~\ref{tab1}). At $\ds=1$\,kpc, the mass distribution would reach a magnitude limit of $V\la31$. 
Then, to emulate photometric depth, we also consider CMDs restricted to $V\le20$ and $V\le17$.

Results obtained by \fC\ can be seen in Fig.~\ref{fig2} and are quantified in Table~\ref{tab1}. The 
first point is that \fC\ does recover the input values - within uncertainties - even for the severely 
depleted CMDs. As expected, the cluster mass determination does not depend on photometric depth and/or
number of stars present in the CMD. Age and metallicity, on the other hand, are more sensitive, especially
to the evolutionary sequence tightness, i.e., essentially the number of stars.

\section{Application to actual stellar systems}
\label{APP}

After exploring the ability of \fC\ to recover input model parameters, we now apply it to 
CMDs of actual star clusters. For this we selected some OCs with photometry publicly available 
in VizieR\footnote{http://vizier.u-strasbg.fr/viz-bin/VizieR}, NGC\,6791, NGC\,188, NGC\,2682, 
NGC\,2635, NGC\,5288, and NGC\,2323. Some details and values of parameters recently derived
for these OCs are provided below.

NGC\,6791: BVR photometry from \citet{Montgomery94}. A recent review on properties on this
old and relatively metal-rich OC is in \citet{Martinez18}. Some relevant parameters are an age 
in the range $6-8$\,Gyr, $\fH\approx+0.4$, $\ds\approx4$\,kpc, and $\mcl\sim5000\ms$.

NGC\,188: BVI photometry from \citet{Sarajedini99}. The review by \citet{Hills15}
provides the following parameter ranges, $\tA=5.8$ to $6.5$\,Gyr, $\fH=-0.77$ to $+0.125$, 
$\mMV=11.441$ to $11.525$, and $A_V=0.162$ to $0.236$.

NGC\,2682 (M\,67): BVR photometry from \citet{Yadav08}. Available parameters in the literature:
$\tA\sim4$\,Gyr, $\mcl\sim2000$\,ms\ (\citealt{Hurley05}), $\fH=0.00\pm0.06$ (\citealt{Heiter14}),
$\EBV=0.041\pm0.004$ (\citealt{Taylor07}).

NGC\,2635: BVI photometry from \citet{Moitinho06}, which also provides the values 
$\tA\sim600$\,Myr, $\EBV=0.35$, $\ds\sim4$\,kpc, and $\mZ\sim0.004$.

NGC\,5288: BVI photometry from \citet{Piatti06}, which also provides the values 
$\tA\sim130$\,Myr, $\EBV=0.75$, $\ds=2.1\pm0.3$\,kpc, and $\mZ\sim0.04$.

NGC\,2323: JHK$_S$ photometry from 2MASS \citet{Skrutskie06}; parameters in the literature
are $\tA=140\pm20$\,Myr, $\ds=115\pm20$\,pc, $\fH=0.00$, and $\EBV=0.23\pm0.06$ 
(\citealt{Cummings16}); $\tA=140\pm20$\,Myr, $\ds=900\pm100$\,pc, and $\mcl\sim890$\ms\ 
(\citealt{Amin17}).

As a caveat, it is important to remark that, except for NGC\,2323, the remaining OCs may 
contain varying fractions of field stars contaminating their CMDs, which may lead to extrinsic 
stellar-density differences between the observed and IMF-simulated evolutionary sequences. 
The wide-field 2MASS photometry of NGC\,2323 (obtained from VizieR) has been field-star 
cleaned with the decontamination algorithm described in \citet{Bonatto07}, which employs a 
comparison field containing a statistically significant number of stars. Photometry for the 
other OCs, on the other hand, was obtained from specific observational projects, and 
usually corresponds to stars located within the cluster radius, thus with no comparison field 
available. Nevertheless, they provide interesting cases in which \fC\ can be tested under 
realistic conditions and those usually found in star clusters.

\begin{figure}
\begin{minipage}[b]{0.5\linewidth}
\includegraphics[width=\textwidth]{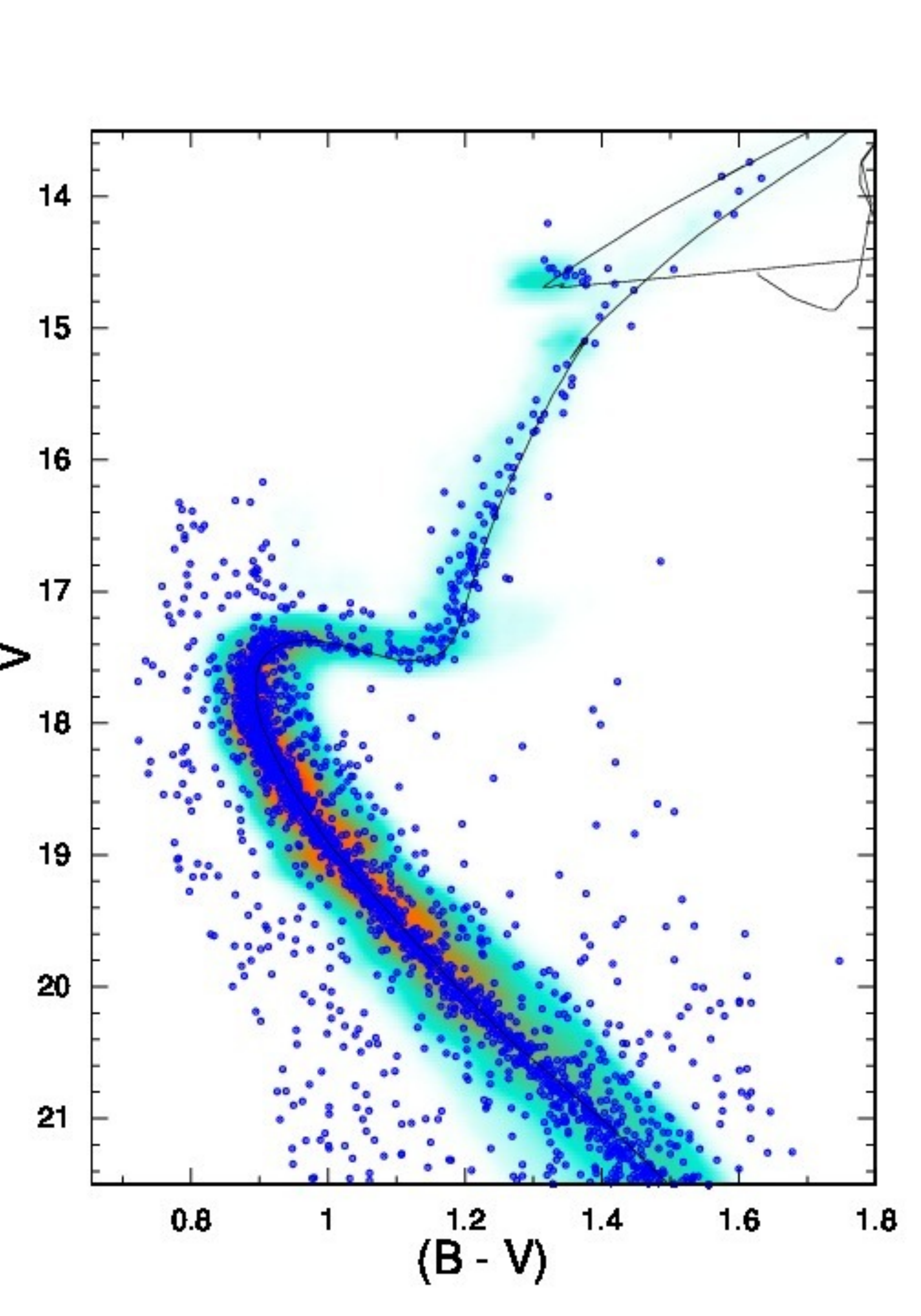}
\end{minipage}\hfill
\begin{minipage}[b]{0.5\linewidth}
\includegraphics[width=\textwidth]{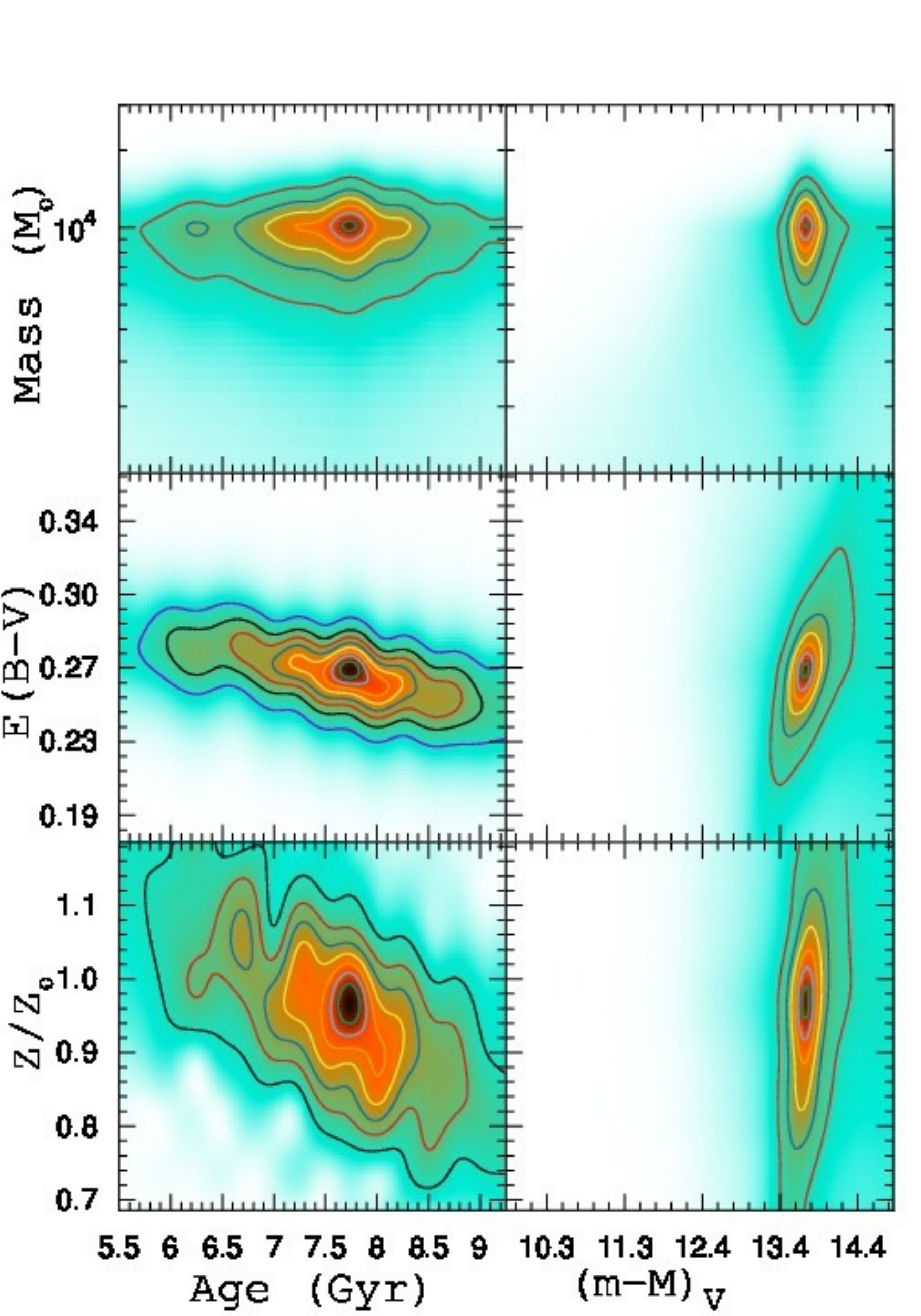}
\end{minipage}\hfill
\begin{minipage}[b]{0.5\linewidth}
\includegraphics[width=\textwidth]{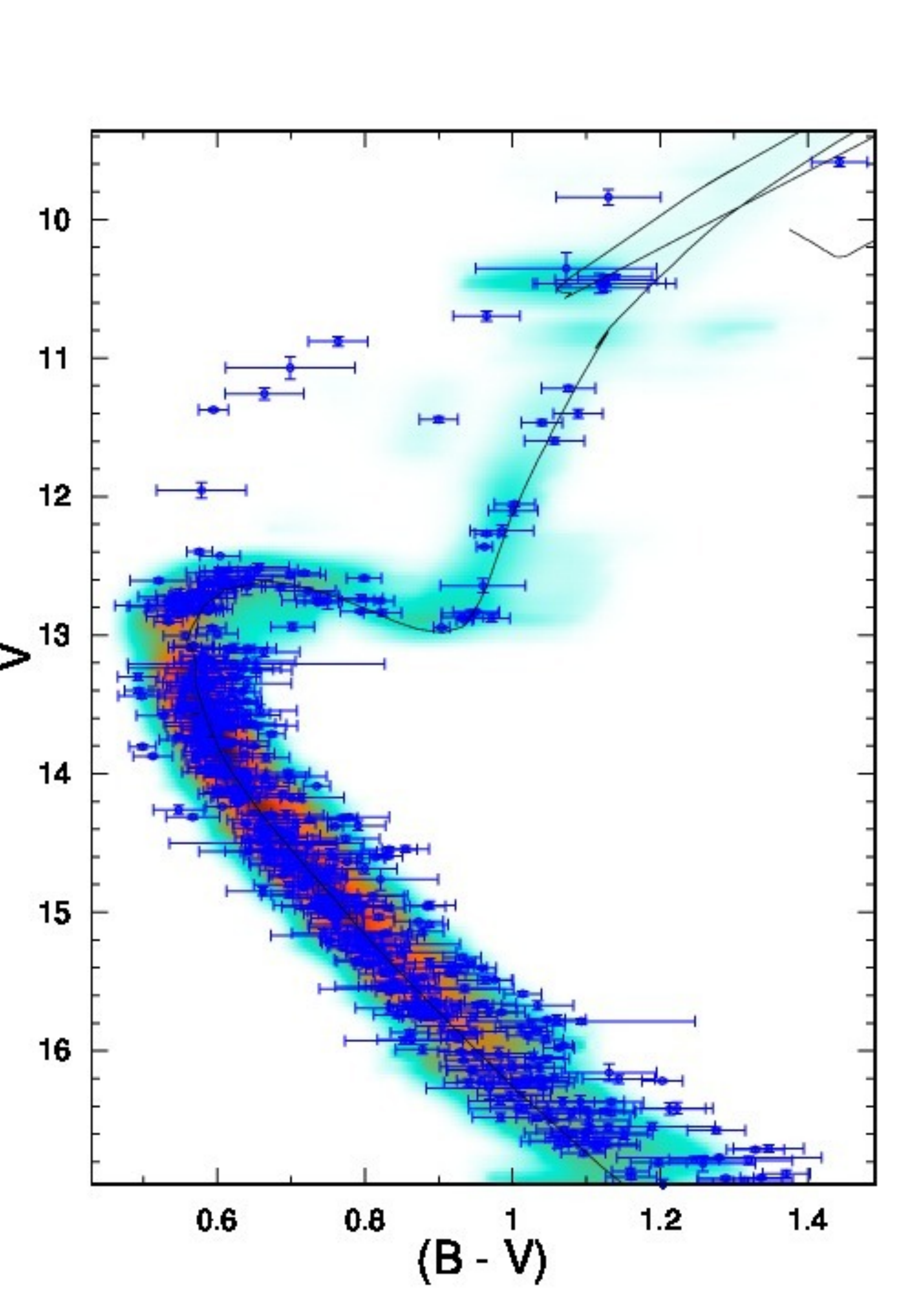}
\end{minipage}\hfill
\begin{minipage}[b]{0.5\linewidth}
\includegraphics[width=\textwidth]{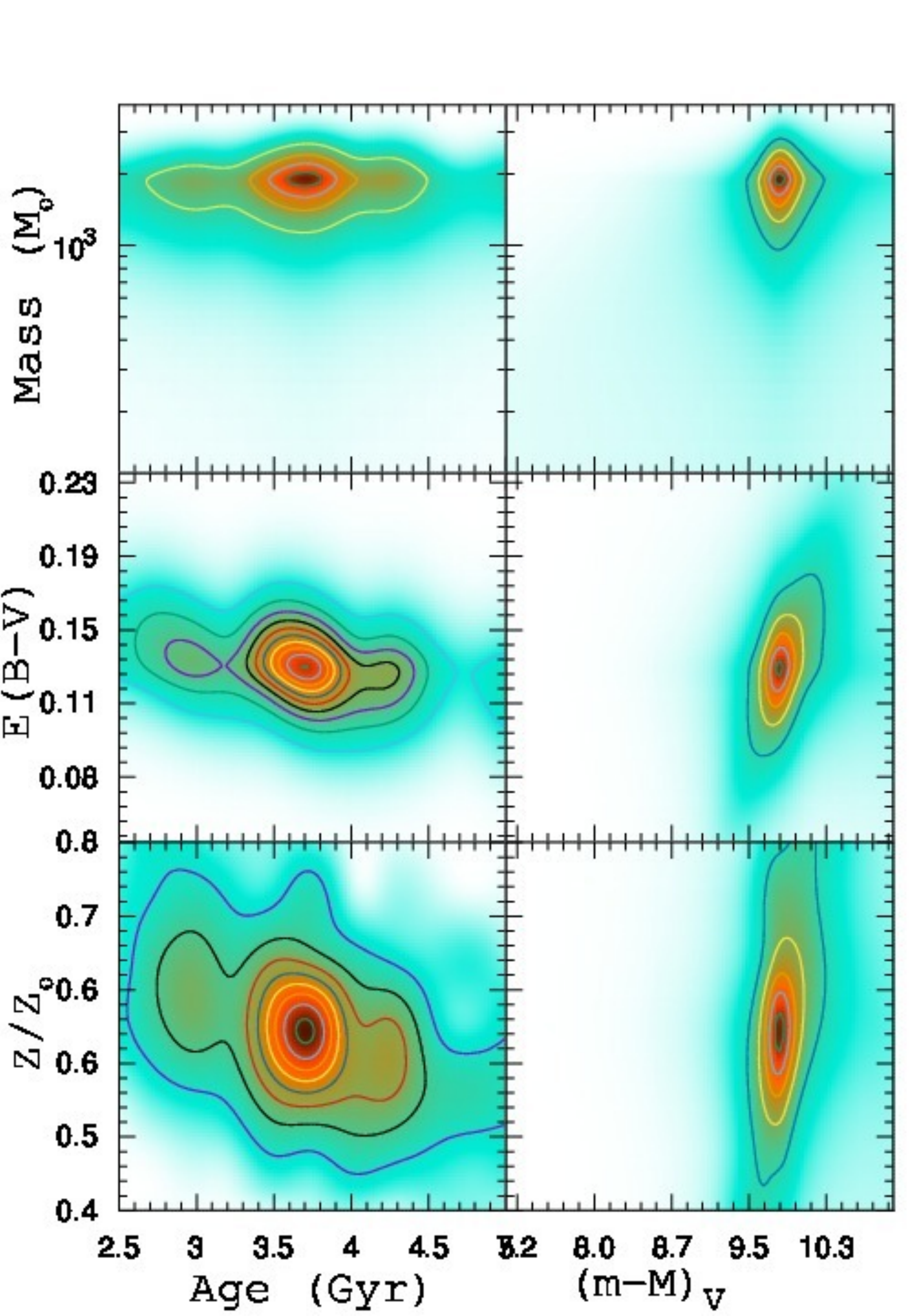}
\end{minipage}\hfill
\caption[]{Left panels: same as Fig.~\ref{fig2} for the old OCs NGC\,6791 (top) and NGC\,2682 
(bottom). The respective solution maps are shown in the right panels.}
\label{fig3}
\end{figure}

\begin{figure}
\begin{minipage}[b]{0.5\linewidth}
\includegraphics[width=\textwidth]{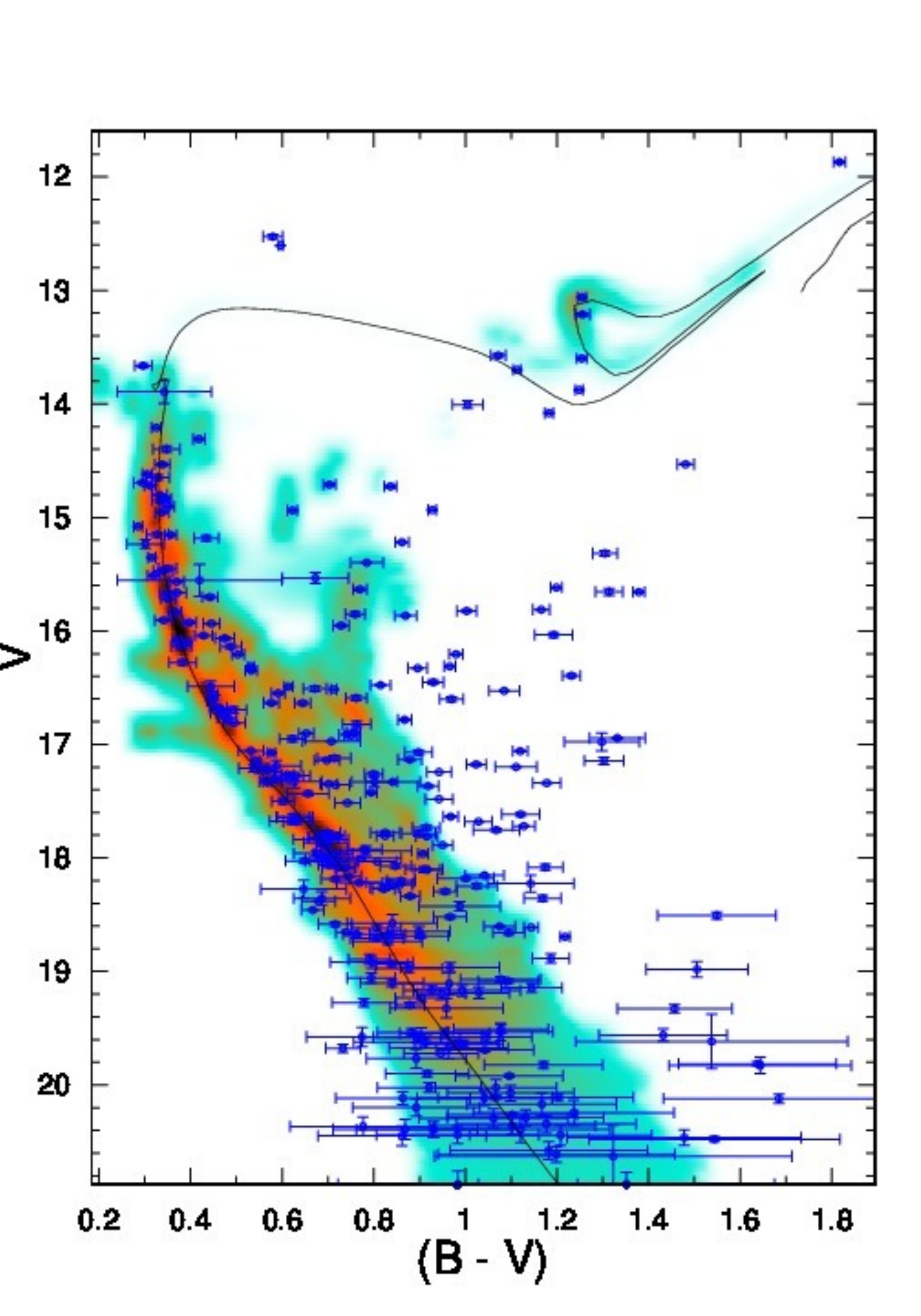}
\end{minipage}\hfill
\begin{minipage}[b]{0.5\linewidth}
\includegraphics[width=\textwidth]{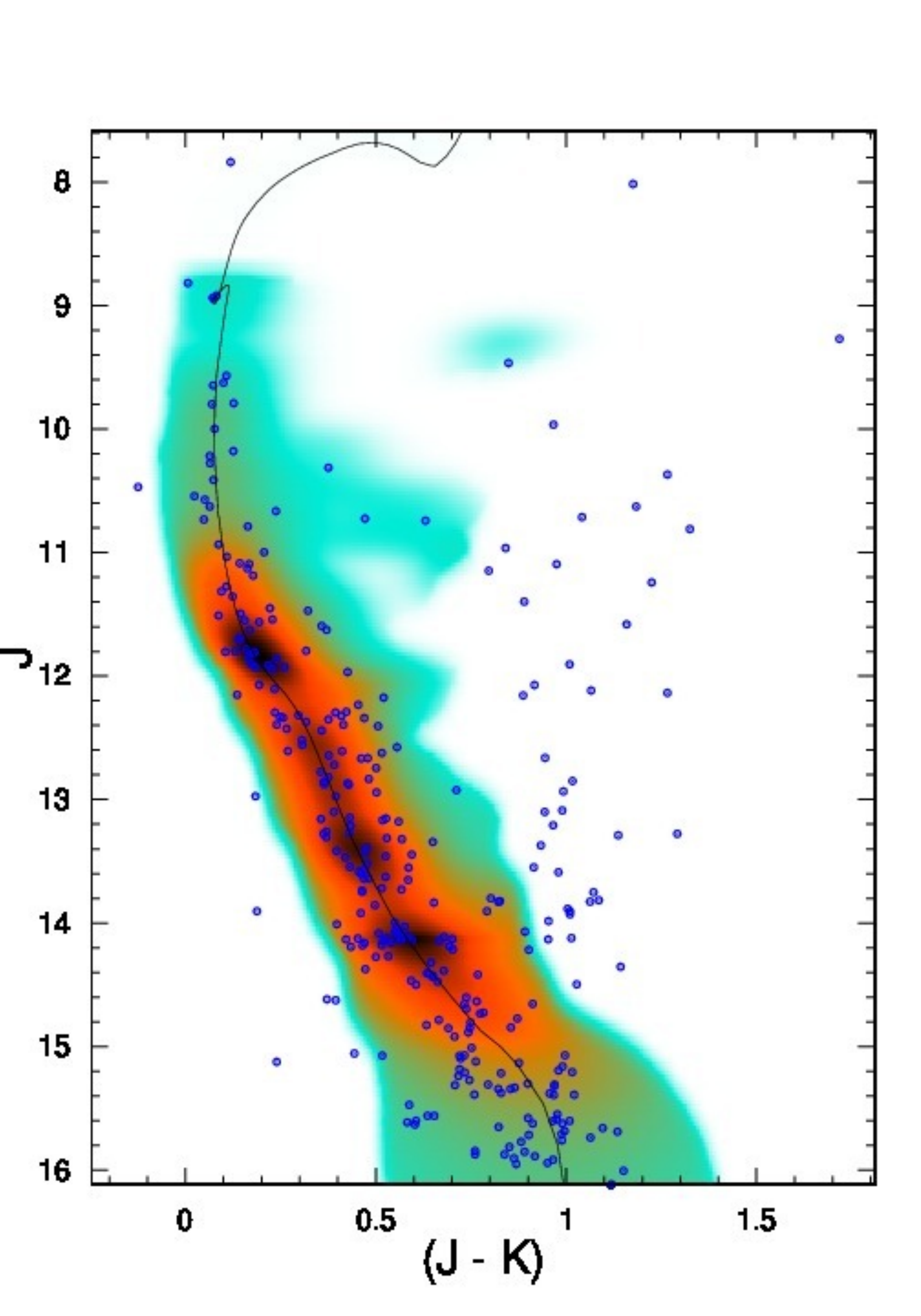}
\end{minipage}\hfill
\begin{minipage}[b]{0.5\linewidth}
\includegraphics[width=\textwidth]{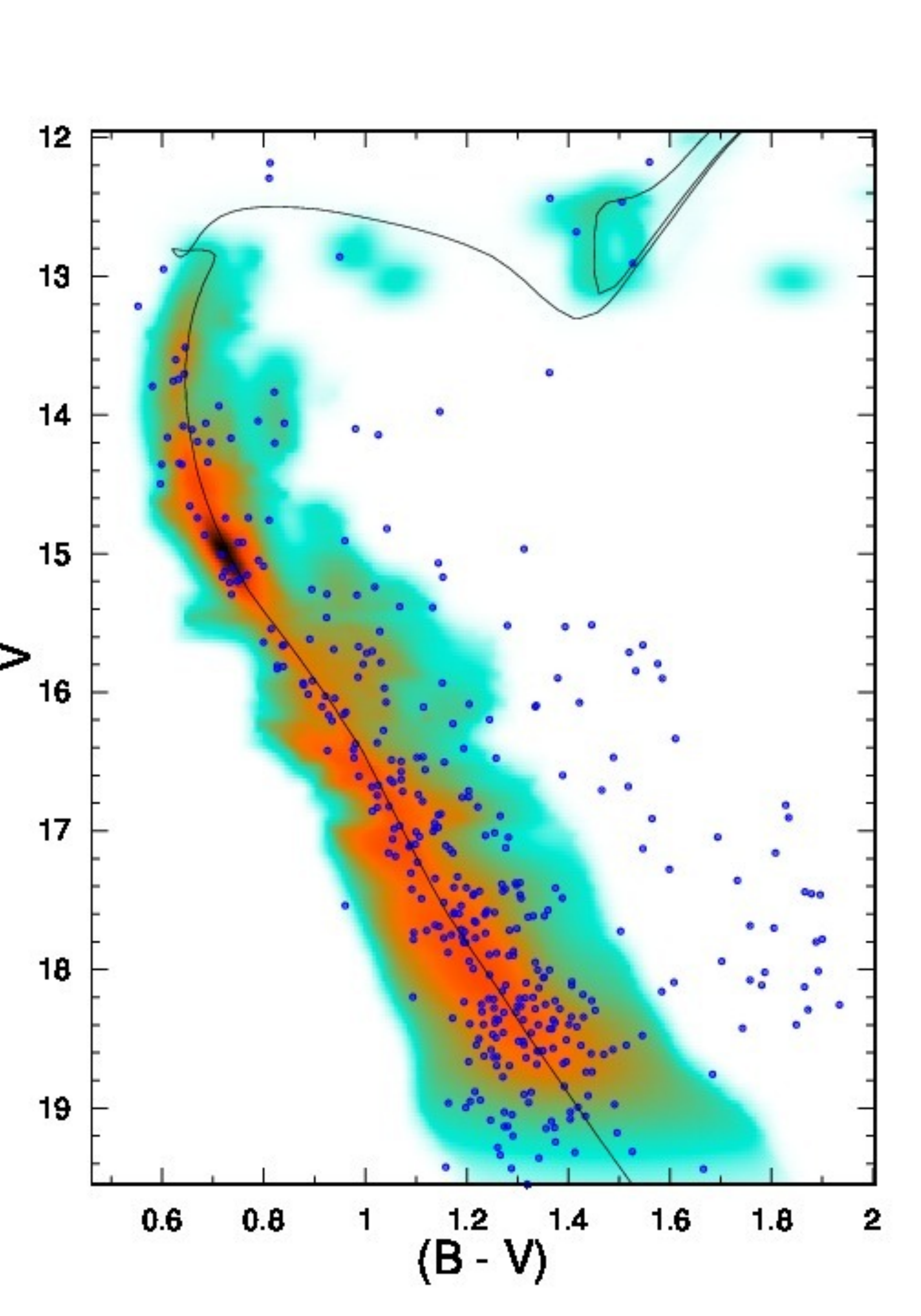}
\end{minipage}\hfill
\begin{minipage}[b]{0.5\linewidth}
\includegraphics[width=\textwidth]{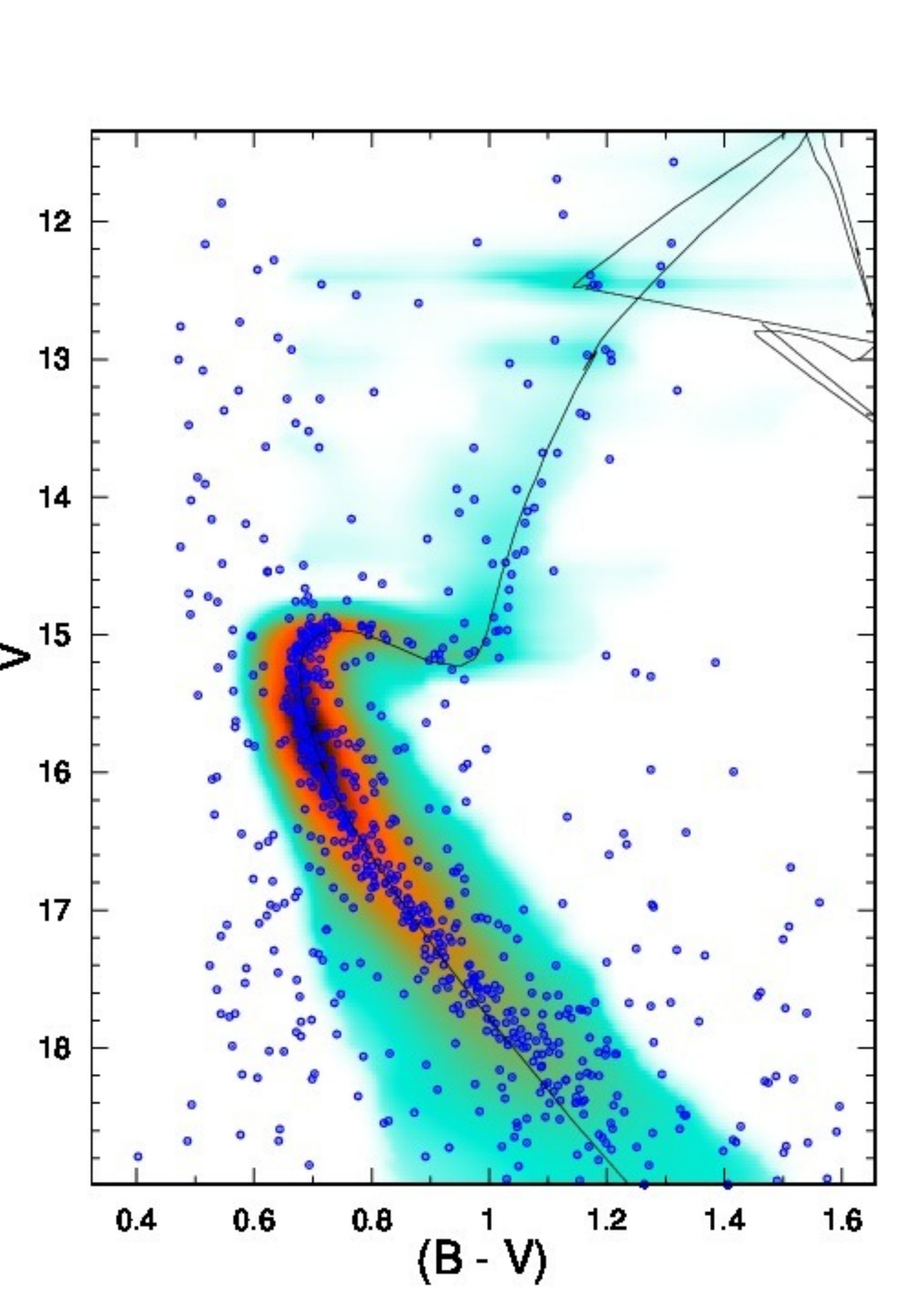}
\end{minipage}\hfill
\caption[]{Same as Fig.~\ref{fig2} for NGC\,2635 (top-left), NGC\,2323 (top-right),
NGC\,5288 (bottom-left), and NGC\,188 (bottom-right).}
\label{fig4}
\end{figure}

In all cases the age and metallicity parameters were allowed to vary over their whole ranges
(Sect.~\ref{TOM}); the remaining search parameters were also let to vary over broad ranges. 
For comparison reasons, simulations not taking photometric completeness into account were also 
performed and will be discussed later. The observed CMDs together with the respective \fC\ 
solutions are shown in Figs.~\ref{fig3}-\ref{fig4}; to minimize clutter, error bars are not shown 
in the cases where the OC has many stars. The best-fit and additional derived parameters are given 
in Table~\ref{tab2}, among them is the stellar mass actually present in the CMD ($M_{CMD}$), which
in some cases corresponds to a fraction of the total mass (\mcl).

\begin{landscape}
\begin{table}
\caption[]{Parameters of selected open clusters derived with \fC}
\label{tab2}
\renewcommand{\tabcolsep}{2.35mm}
\begin{tabular}{cccccccccccccc}
\hline\hline
Cluster&\mcl&age&\mMV  &\EBV &\Zo&$M_{CMD}$&\ds  &$M_{bol}$&$M_V$&MLR$_{bol}$&MLR$_V$&\qA&\qB \\
      &($10^3\ms$)&(Gyr)&(mag)&(mag)&   &  (\ms)  &(kpc)&(mag)&(mag)&$(\ms/\ls)$&$(\ms/\ls)$ & &(mag)\\
(1)&(2)&(3)&(4)&(5)&(6)&(7)&(8)&(9)&(10)&(11)&(12)&(13)&(14)\\
\hline
NGC\,6791 &$10.2\pm1.4$&$7.7\pm0.4$&$13.74^{+0.24}_{-0.08}$&$0.27\pm0.01$&$1.00\pm0.05$&1700&$3.74^{+0.41}_{-0.14}$&-5.0&-4.2&1.33&2.50&$1.02\pm0.15$&$20.2\pm0.4$ \\

NGC\,188&$4.7\pm0.5$&$6.3\pm0.2$&$11.41^{+0.12}_{-0.09}$&$0.02\pm0.01$&$1.58\pm0.09$&520&$1.86^{+0.10}_{-0.07}$&-4.2&-3.3&1.25&2.58&$0.65\pm0.10$&$16.4\pm0.3$\\

NGC\,2682&$1.9\pm0.3$&$3.8\pm0.2$&$9.87^{+0.23}_{-0.08}$&$0.13\pm0.01$&$0.59\pm0.04$&390&$0.78^{+0.08}_{-0.03}$&-3.8&-3.2&0.70&1.18&$1.06\pm0.16$&$15.8\pm0.3$ \\

NGC\,2635&$1.0\pm0.2$&$0.28\pm0.03$&$14.77^{+0.82}_{-0.38}$&$0.43\pm0.02$&$0.53\pm0.05$&330&$4.72^{+1.77}_{-0.83}$&-4.9&-4.4&0.13&0.20&$1.18\pm0.18$&$19.6\pm0.5$ \\

NGC\,5288&$0.8\pm0.1$&$0.55\pm0.04$&$13.05^{+0.49}_{-0.28}$&$0.67\pm0.03$&$0.46\pm0.05$&310&$1.48^{+0.33}_{-0.19}$&-4.1&-3.8&0.23&0.29&$6.40\pm0.98$&$19.1\pm0.5$ \\

NGC\,2323&$0.6\pm0.1$&$0.32\pm0.03$&$9.74^{+0.21}_{-0.16}$&$0.21\pm0.03$&$0.99\pm0.14$&270&$0.81^{+0.08}_{-0.06}$&-4.1&-5.6&0.17&0.23&$1.42\pm0.22$&$15.2\pm0.3$\\
\hline
\end{tabular}
\begin{list}{Table Notes.}
\item This table also gives the bolometric (col.~9) and absolute V (col.~10) magnitudes, the bolometric 
(col.~11) and V (col.~12) mass to light ratios; Fermi function parameters related to photometric completeness
are in cols.~(13) and (14).  
\end{list}
\end{table}
\end{landscape}

After finding the parameters corresponding to the absolute minimum of the \x2\ surface, their
uncertainties are computed by means of the shape of the solution well in selected 2D projections 
(or solution maps). For instance, the solution map for mass and age is built by setting \DM, 
\mZ, \CE, \qA, and \qB\ to their optimum values, while keeping \mcl\ and \tA\ free to vary within 
the pre-defined ranges. Uncertainties - usually asymmetric - are then computed based on properties 
(depth and width of the absolute minimum) of the solution map. Examples of solution maps are 
given in Fig.~\ref{fig3} for NGC\,6791 and NGC\,2682.

Besides the magnitude values as a function of mass, the PARSEC isochrones provide additional
data that allow us to compute interesting star cluster integrated quantities, such as the bolometric 
and the absolute magnitude in the V band, and the mass-to-light ratio (MLR) both bolometric and in V
(Tab.~\ref{tab2}). 

A few comments on the results obtained with \fC\ for the selected OCs. A direct comparison of all 
the output parameters from \fC\ with those in the literature cannot be done, since usually each
work is aimed at a particular parameter (or restricted set of parameters). Nevertheless, the \fC\ 
ages agree - within the uncertainties - with those in the literature for NGC\,6791, NGC\,188 and 
NGC\,2682, but differ somewhat for the remaining OCs. Concerning \mcl, our values for NGC\,2682 and 
NGC\,2323 agrees with previous ones, but we find a value twice as large for NGC\,6791. In summary, 
\fC\ provides values for a set of interesting astrophysical parameters obtained in a self-consistent 
way.

\begin{figure}
\resizebox{\hsize}{!}{\includegraphics[width=\textwidth]{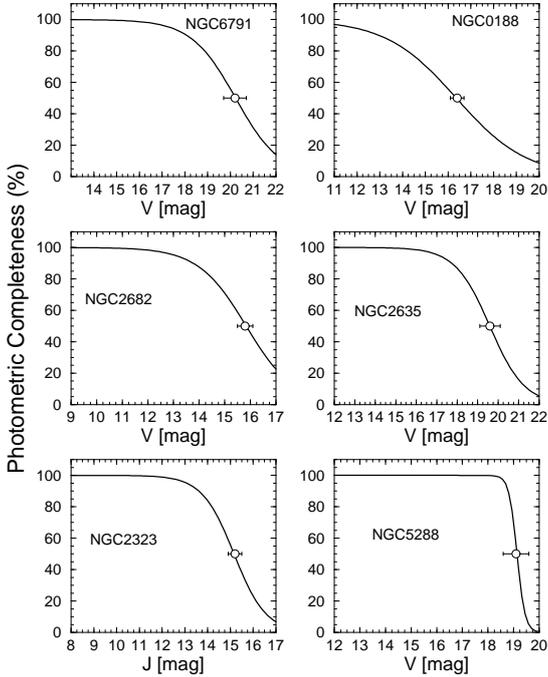}}
\caption[]{Photometric completeness ($f_C$) as a function of magnitude ($m$): 
$f_C(m) = 1/(1+\exp{[\qA(m-\qB)]})$. Turnover magnitude (\qB) is shown by the open circle.}
\label{fig5}
\end{figure}

The 2 free parameters related to the photometric completeness are given in Tab.~\ref{tab2}, 
and the completeness functions derived for the sample OCs are shown in Fig.~\ref{fig5}. Among
the selected OCs, NGC\,188 is the one that suffers the highest degree of photometric completeness,
beginning even at the brightest observed magnitude ($V\sim11$ and $\qB\approx16.4$). NGC\,5288, on 
the other hand, is the least affected, with completeness only affecting stars fainter than $V\sim18$, 
with $\qB\approx19.1$.

\begin{figure}
\resizebox{\hsize}{!}{\includegraphics[width=\textwidth]{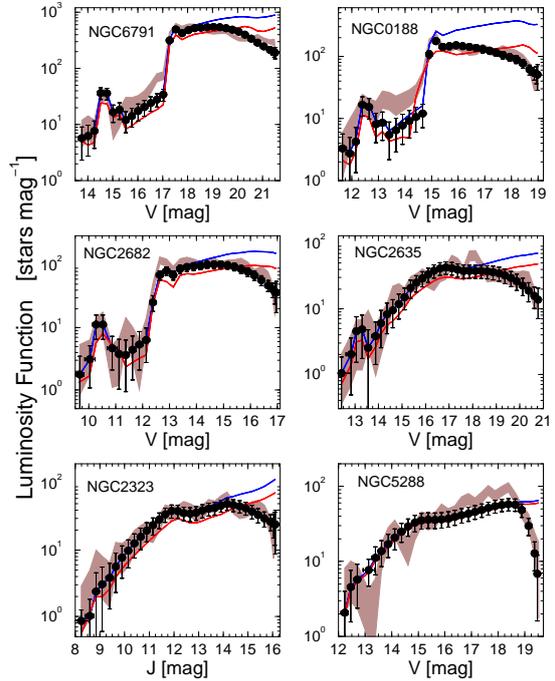}}
\caption[]{Luminosity functions. Shaded region: $1\,\sigma$ bounds of the observed LF;
black symbols: \fC\ simulated LF; blue solid line: completeness-corrected LF; red line: 
best-fit LF without considering completeness. }
\label{fig6}
\end{figure}

Another way to examine the results is by comparing the observed luminosity function (LF) with that 
measured on the CMD built with the best-fit parameters (Fig.~\ref{fig6}). The LFs are built by simply 
integrating the stellar density across the magnitude axis, thus implying that some contribution from
contaminant stars may be present in the observed LFs. For a cleaner visualization, Fig.~\ref{fig6} 
shows the $1\,\sigma$ bounds of the observed LFs. In general, \fC\ reproduces the observed LFs, within 
the uncertainties, even at the faint end where completeness is more important. For comparison purposes,
Fig.~\ref{fig6} also shows the LF produced by \fC\ when photometric completeness is not taken into account 
(red line). Significant deviations occur with respect to both the complete and observed LFs, especially 
at the faint magnitude ranges for the OCs most affected by completeness. This occurs because \fC\ tries 
to match the observed stellar density in all cells according to an IMF. Since photometric completeness
naturally decreases the stellar densities (especially at the faint end of CMDs), \fC\ also lowers the 
stellar density in brighter cells. Finally, Fig.~\ref{fig6} shows the completeness-corrected LFs (blue 
line).

\subsection{Special cases}
\label{specialCases}

After being applied to artificial and classical OCs, it would be interesting to check the performance
of \fC\ on potentially more difficult cases. To this end, the Milky Way satellite Reticulum\,II and 
the globular cluster (GC) 47\,Tuc have been selected. Both objects contain mixtures of stellar 
populations (see below) probably characterized by different ages and metallicities. 

The dwarf galaxy Ret\,II (DES\,J0335.6-5403) was discovered by the Dark Energy Survey (DES; \citealt{Bechtol2015}). 
Parameters for Ret\,II derived by \citet{Bechtol2015} are the half-light radius $R_{hl}\approx6'$, distance 
to the Sun $\ds=32$\,kpc, stellar mass $\mcl=(2.6\pm0.2)\times10^3\,\ms$, absolute magnitude $M_V=-3.6\pm0.1$,
$age=10\pm5$\,Gyr, and the global metallicity $Z<0.0003$. More recently, \citet{Mutlu2018} studied Ret\,II 
with deep Magellan/Megacam photometry (also for stars within $R_{hl}$), finding $\ds=31.4\pm1.4$\,kpc, 
$M_V=-3.1\pm0.1$, $age=13.5$\,Gyr, and $[Fe/H]=-2.4$; for $0.0<[\alpha/Fe]<0.4$ , the total metallicity of 
Ret\,II would be $Z<0.0001$, consistent with the value derived by \citet{Bechtol2015}.

\begin{figure}
\begin{minipage}[b]{0.5\linewidth}
\includegraphics[width=\textwidth]{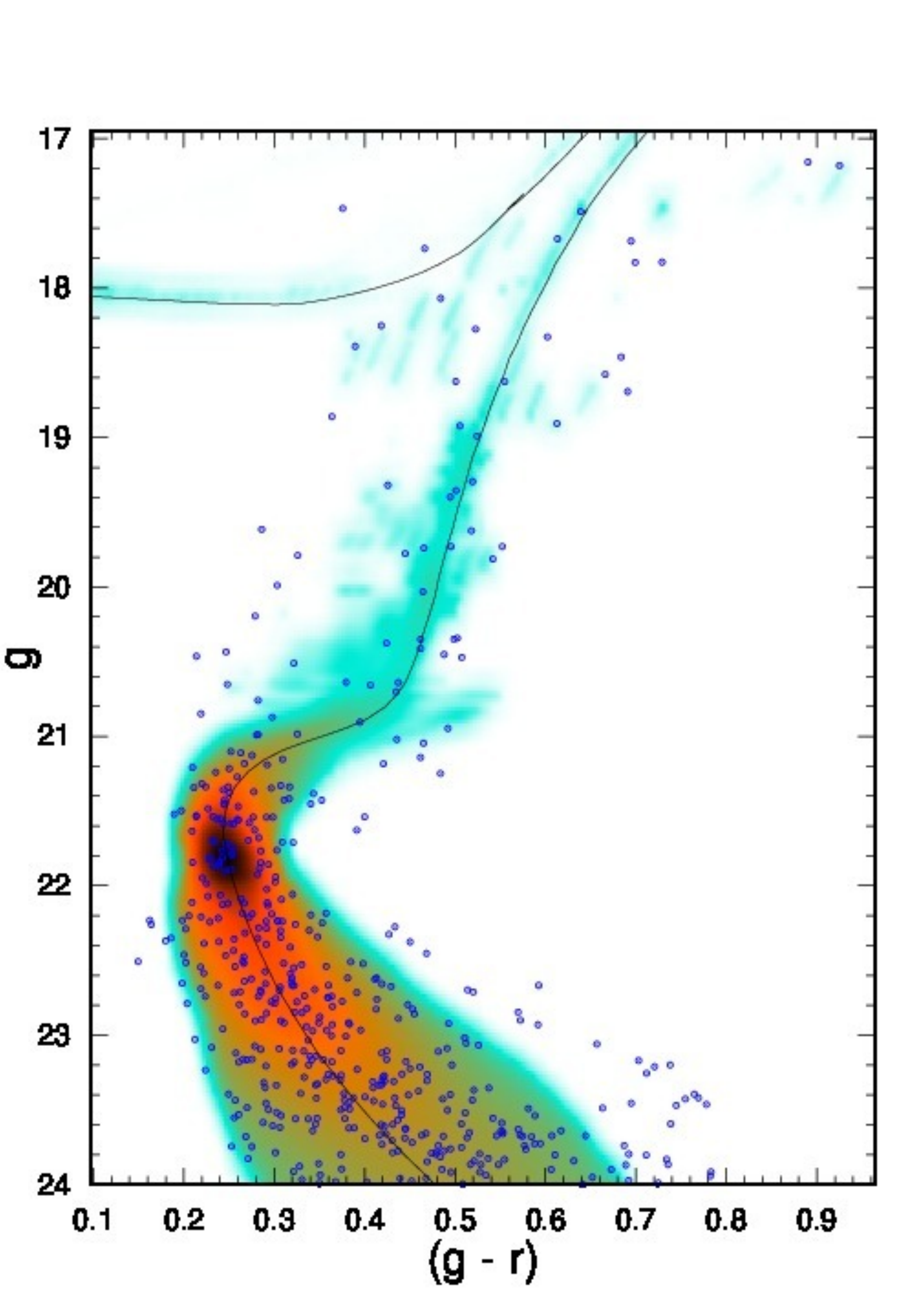}
\end{minipage}\hfill
\begin{minipage}[b]{0.5\linewidth}
\includegraphics[width=\textwidth]{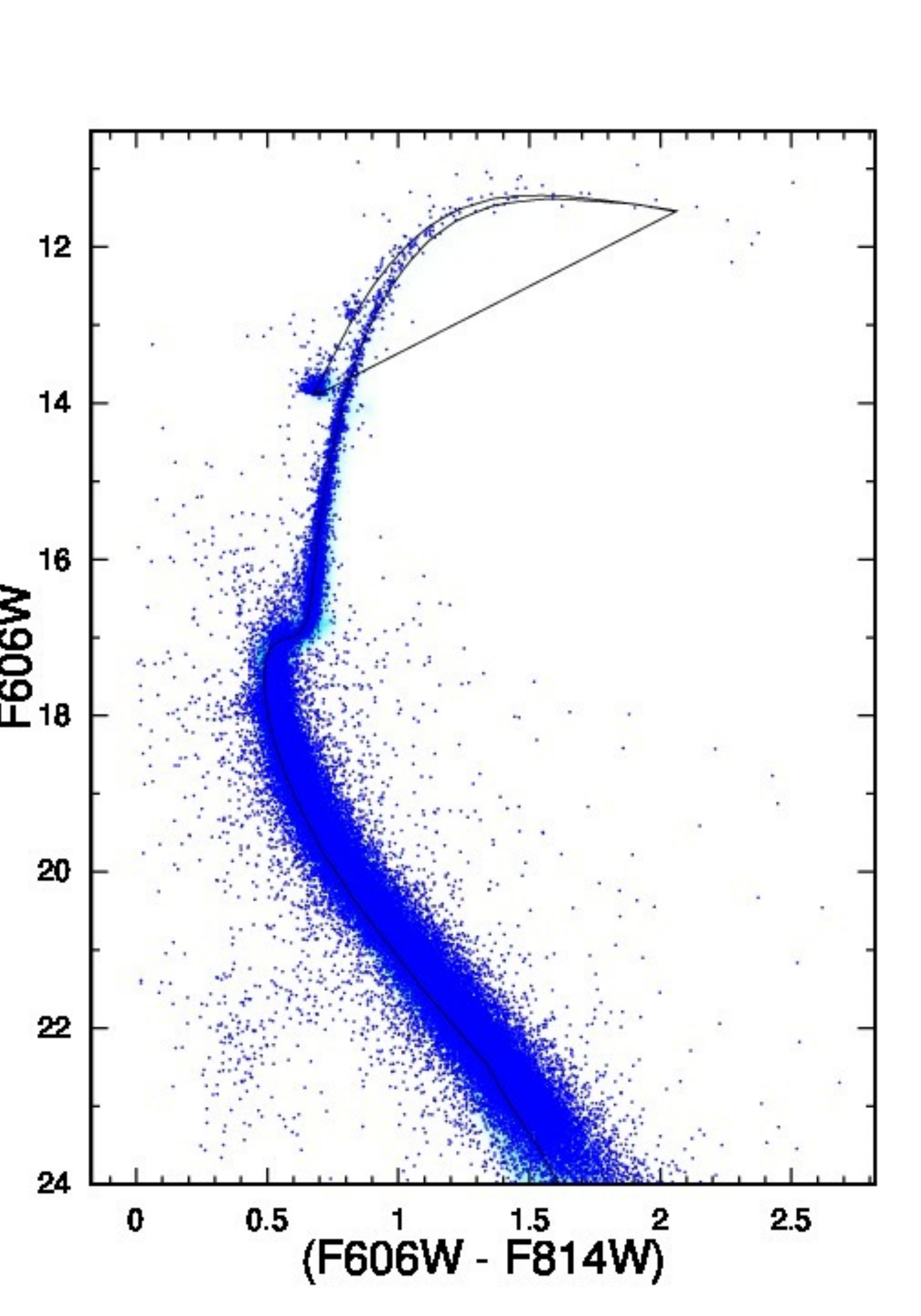}
\end{minipage}\hfill
\caption[]{Left panels: same as Fig.~\ref{fig2} for the dwarf galaxy Ret\,II (left) and the 
GC 47\,Tuc (right). }
\label{fig7}
\end{figure}

\begin{figure}
\resizebox{\hsize}{!}{\includegraphics[width=\textwidth]{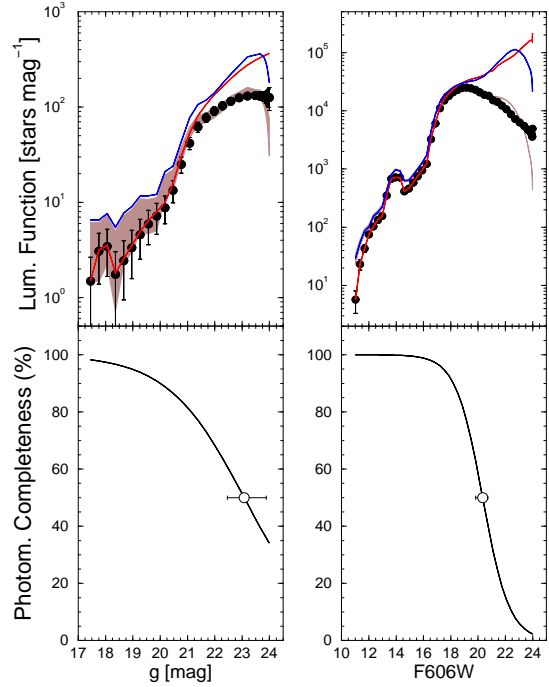}}
\caption[]{Top: same as Fig.~\ref{fig6} for Ret\,II (left) and 47\,Tuc (right); bottom:
same as Fig.~\ref{fig5}.}
\label{fig8}
\end{figure}

Photometry for Ret\,II stars in $g$ and $r$ bands (corresponding to the DES system) has been obtained 
from the NOAO Data Lab\footnote{https://datalab.noao.edu/query.php?name=des$\_$dr1.main} in a region of 
$30'$ radius around its central coordinates, $RA(J2000)=03^h\,35'\,49''$ and $DEC(J2000)=-54\degr\,02'\,48''$. 
This setup was necessary in order to produce a field-stars decontaminated CMD by means of the 
\citet{Bonatto07} algorithm. For consistency and comparison purposes with both previous works, the region
analyzed here corresponds to the area within $R_{hl}$. The decontaminated CMD together with the \fC\ solution 
are shown in Fig.~\ref{fig7}. The corresponding \fC\ parameters are $\mcl=(2.7\pm0.5)\times10^3\,\ms$ (only 
$\approx8\%$ of this mass is present on the observed CMD), $age=13.0^{+0.3}_{-0.9}$\,Gyr, $Z=0.0002$, 
$\ds=30.6\pm1.4$\,kpc, the absolute magnitude in $g$ $M_g=-3.2$, and the bolometric magnitude $M_{bol}=-3.72$. 
V magnitudes can be obtained from $g$ and $r$ by the transformation $V=g-0.487(g-r)-0.025$ (\citealt{Bechtol2015}), 
resulting in $M_V=-3.5$. Within the quoted uncertainties, these parameters agree with the previous ones, except 
for the lower metallicity implied by \citet{Mutlu2018}. Luminosity and photometric completeness functions are 
shown in Fig.~\ref{fig8}. Consistently with ground-based observations of a distant object, Ret\,II 
photometry appears to be affected by completeness issues even at the bright end of the LF.

47\,Tucanae (NGC\,104) is the second brightest Milky Way GC after {\em Omega Centauri} (NGC\,5139). 
Its tens of thousands stars distribute across $\sim50'$ of the southern sky. 47\,Tuc lies at 
$\ds\approx4.5$\,kpc from the Sun, displays a small and dense central core ($R_c\approx0'.36$) 
and has half-light and tidal radii of $R_{hl}\approx3.2'$ and $r_t\approx43'$, respectively 
(\citealt{Harris2010}). More recently, 47\,Tuc was shown to host multiple stellar populations 
(\citealt{Milone2012}). 

High-quality photometry for 47\,Tuc is available as part of the HST WFC/ACS GC sample under program 
number GO\,10775, with A. Sarajedini as PI. GO\,10775 is a HST Treasury project in which 66 GCs were 
observed through the F606W and F814W filters with a field of view of 
$\approx200\arcsec\times200\arcsec$ (\citealt{Sarajedini2007}). Working with this photometry, 
\citet{WKaiser2017} found $A_V=0.105\pm0.002$, 
$[Fe/H]=-0.72$ ($[\alpha/Fe]\approx+0.4$), which corresponds to $Z\approx0.006$, and the rather 
old value $age=13.494^{+0.006}_{-0.022}$\,Gyr. On the other hand, working with deep WFC3 IR 
photometry, \citet{Correnti2016} derived $age=11.6\pm0.7$\,Gyr. It is clear that there is no 
consensus on the exact age of 47\,Tuc. Indeed, \citet{Brogaard2017} find the value 11.8\,Gyr, 
with lower and upper limits ($3\,\sigma$) at 10.4 and 13.4\,Gyr, a range that appears to 
realistically represent the actual age uncertainty of 47\,Tuc. Concerning the stellar mass of
47\,Tuc, N-body computations of \citet{MarksKroupa10} provide $7\times10^5\,\ms$, while analysis
of central velocity dispersions coupled to fits of dynamical models by \citet{Kimmig15} imply 
$(7\pm1)\times10^5\,\ms$.

The \fC\ analysis of 47\,Tuc is based on the \citet{Sarajedini2007} photometry; as an additional 
quality constraint, only stars with photometric uncertainty lower than 0.1\,mag in F606W and F814W 
(reducing to $\approx128000$ the number of available stars) are considered. The \fC\ solution for 
47\,Tuc is shown in Fig.~\ref{fig7}, and the best-fit parameters are $\mcl=(3.4\pm0.4)\times10^5\,\ms$ 
($\approx24\%$ of this is on the observed CMD), $age=12.0^{+0.3}_{-0.5}$\,Gyr, $Z=0.004$, 
$\ds=4.5\pm0.1$\,kpc, the absolute magnitude in $F606W$ $M_{F606W}=-8.34$, the bolometric 
magnitude $M_{bol}=-8.63$, and $A_V=0.11\pm0.01$. \fC\ age agrees quite well with the values 
of \citet{Brogaard2017} and \citet{Correnti2016}. The \fC\ value for the mass corresponds to 
about half of the dynamical estimates (see above). However, it should be noted that the \fC\ 
value is based on the stars present in the area sampled by WFC/ACS, which has a radius corresponding 
to about half of $R_{hl}$. Interestingly, photometric completeness is inconspicuous for $F606W\la16$, 
but falls off quite steeply for $F606W\ga18$ (Fig.~\ref{fig8}).

\section{Concluding remarks}
\label{CONC}

\fC\ is an approach that extracts a set of astrophysical parameters from CMDs of star clusters. 
The rationale is to transpose theoretical IMF properties to their observational counterpart, the 
CMD. This requires finding values of the total (or cluster) stellar mass, age, global metallicity, 
foreground reddening, distance modulus, as well as for parameters describing magnitude-dependent 
photometric completeness. These parameters - including photometric scatter - are used to build 
a synthetic CMD that is compared with that of a star cluster. Residual minimization between observed 
and synthetic CMDs - by means of the global optimization algorithm Simulated Annealing - then leads 
to the best-fit parameters.

The efficiency of \fC, both in terms of computational time and ability to recover input parameters,
has been tested with CMDs of artificial and observed star cluster - as well as of a dwarf galaxy 
and a globular cluster, with excellent results. In principle, \fC\ can be used with any isochrone set 
that provides magnitudes for at least 2 different bands for stellar masses covering as wide as possible 
a range. In addition, the isochrone set should also provide a comprehensive coverage - and resolution - 
both in age and metallicity. 

\section*{Acknowledgements}
Thanks to an anonymous referee for important comments and suggestions.
This research has made use of the VizieR catalogue access tool, CDS, Strasbourg, France. 
This publication makes use of data products from the Two Micron All Sky Survey, which is a joint 
project of the University of Massachusetts and the Infrared Processing and Analysis Center/California
Institute of Technology, funded by the National Aeronautics and Space Administration and the National 
Science Foundation. C.B. acknowledges support from the Brazilian Institution CNPq.


\bibliographystyle{mnras}
\bibliography{ref}

\end{document}